\documentclass[twocolumn]{aastex63}

\usepackage{amsmath}	
\usepackage{amssymb}	

\newcommand{\teff}{$T_{\text{eff}}$}%
\newcommand{\fkep}{$F_{Kep}$}%
\newcommand{\kepler}{\textit{Kepler}}%
\newcommand{\gaia}{\textit{Gaia}}%
\newcommand{\npbar}{$\bar{N}_p$}

\begin{document}

\title{Occurrence and Architecture of \kepler\ Planetary Systems as Functions of Stellar Mass and Effective Temperature}

\correspondingauthor{Ji-Wei Xie}
\email{jwxie@nju.edu.cn}


\author{Jia-Yi Yang}
\affil{School of Astronomy and Space Science, Nanjing University, Nanjing 210023, People's Republic of China}
\affil{Key Laboratory of Modern Astronomy and Astrophysics, Ministry of Education, Nanjing 210023, People's Republic of China}

\author{Ji-Wei Xie}
\affil{School of Astronomy and Space Science, Nanjing University, Nanjing 210023, People's Republic of China}
\affil{Key Laboratory of Modern Astronomy and Astrophysics, Ministry of Education, Nanjing 210023, People's Republic of China}

\author{Ji-Lin Zhou}
\affil{School of Astronomy and Space Science, Nanjing University, Nanjing 210023, People's Republic of China}
\affil{Key Laboratory of Modern Astronomy and Astrophysics, Ministry of Education, Nanjing 210023, People's Republic of China}

\begin{abstract}
The \kepler\ mission has discovered thousands of exoplanets around various stars with different spectral types (M, K, G, and F) and thus different masses and effective temperatures. 
Previous studies have shown that the planet occurrence rate, in terms of the average number of planets per star, drops with increasing stellar effective temperature (\teff).
In this paper, with the final \kepler\ Data Release (DR25) catalog, we revisit the relation between stellar effective temperature (as well as mass) and planet occurrence, but in terms of the fraction of stars with planets and the number of planets per planetary system (i.e., planet multiplicity). 
We find that both the fraction of stars with planets and planet multiplicity decrease with increasing stellar temperature and mass. 
Specifically, about 75\% late-type stars (\teff $<$ 5000 K) have \kepler-like planets with an average planet multiplicity of $\sim$2.8,
while for early-type stars (\teff $>$ 6500 K), this fraction and the average multiplicity fall down to $\sim$35\% and $\sim$1.8, respectively. 
The decreasing trend in the fraction of stars with planets is very significant with $\Delta$AIC$>30$, though the trend in planet multiplicity is somewhat tentative with $\Delta$AIC$\sim5$.
Our results also allow us to derive the dispersion of planetary orbital inclinations in relationship with stellar effective temperature.
Interestingly, it is found to be similar to the well-known trend between obliquity and stellar temperature, indicating that the two trends might have a common origin.

\end{abstract}


\section{Introduction}
With the discovery of thousands of exoplanets, the \kepler\ mission \citep{Borucki_2010_Sci_327_977B} provides us an unprecedented sample to study exoplanets statistically. The bulk of \kepler\ planets are the so-called super-Earths or sub-Neptunes with radii between Earth and Neptune and orbital periods within a few hundred days. One of the fundamental questions is: how common are these \kepler\ planetary systems? 

This question can be addressed from two angles of view: the average number of planets per star ($\eta_p$) and the fraction of stars with planets ($F_p$). These two occurrence rates are different but related to each other through 
\begin{equation}\label{eq:fpnp}
	\eta_p = F_p * \bar{N}_p,
\end{equation}
 where \npbar\ is the average planet multiplicity, i.e., the average number of planets per system with planets.    

The average number of planets per star ($\eta_p$) is relatively straightforward to derive from the \kepler\ data. 
Previous studies have shown that planets are generally common with $\eta_p\sim1$ \citep{Howard_2012_ApJS_201_15H, Dong.2013ApJ...778...53D, Fressin_2013_ApJ_766_81F, Batalha.2014PNAS..11112647B,Burke.2015ApJ...809....8B}, though the specific numbers for different types of planets orbiting different types of stars differ significantly \citep{Foreman-Mackey.2014ApJ...795...64F,Dressing_2015_ApJ_807_45D,Mulders_2015_ApJ_798_112M,Silburt.2015ApJ...799..180S,Mulders_2018_haex_bookE_153M, Narang_2018_AJ_156_221N,Hsu.2019AJ....158..109H,Hardegree-Ullman.2019AJ....158...75H}.
Nevertheless, to derive the fraction of stars with planets ($F_p$), one needs additional assumptions or constraints on the intrinsic architecture of planetary systems.
By assuming planets in multi-planet systems are in coplanar orbits,  \citet{Fressin_2013_ApJ_766_81F} and \citet{Petigura_2013_PNAS_11019273P} have found over 50\% of Sun-like stars have \kepler-like planets ($F_p>0.5$).
Recently,  \citet{Zhu_2018_ApJ_860_101Z}, \citet{Mulders.2018AJ....156...24M}, and \citet{He.2019MNRAS.490.4575H} have modified the $F_p$ estimate by taking into account  non-coplanar planetary systems. 

An important step forward is to link the occurrence and architecture of planetary systems to the properties of their hosts, which may shed light on how planets form and evolve around various kinds of stars.
One of the key factors is stellar mass (or effective temperature equivalently for main-sequence stars). 
It has been well established that the properties and lifetimes of protoplanetary disks depend on the masses of their host stars \citep{Williams_2011_ARA&A_49_67W,Barenfeld_2016_ApJ_827_142B,Pascucci_2016_ApJ_831_125P}. 
Since planets are born in disks, thus it is expected that stellar mass plays a crucial role in planet formation and in shaping the final planetary system architecture \citep{Ida_2004_ApJ_604_388I,Alibert.2011A&A...526A..63A}. 
For giant planets, e.g., Jovian planets, using radial velocity survey data, \citet{Johnson_2010_PASP_122_905J} and \citet{Ghezzi.2018ApJ...860..109G} found that the occurrence rate generally scales linearly with stellar mass. 
For smaller planets, e.g., the bulk of planets found by the \kepler\ mission, \citet{Howard_2012_ApJS_201_15H} and \citet{Mulders_2015_ApJ_798_112M} found that the occurrence rate (in terms of the average number of planets per star, $\eta_p$) is anticorrelated to the stellar effective temperature.

In this paper, we revisit the relation between host stellar effective temperature (as well as mass) and planet occurrence rate from the other angle of view, i.e., the fraction of stars with planets ($F_p$).
One of the advantages of using $F_p$ instead of $\eta_p$ is that it is inevitable to derive the average planet multiplicity, $\bar{N}_p$, during the derivation of $F_p$ \citep{Zhu_2018_ApJ_860_101Z}. 
Since planetary systems of different multiplicities may have different orbital properties, e.g., orbital eccentricity and mutual inclination \citep{Xie_2016_PNAS_11311431X}, thus we are able to investigate the effects of stellar properties (e.g., \teff) on not just occurrence but also the architecture and orbital configuration of planetary systems.

This paper is organized as the follows. In Section \ref{sec:datasample} and \ref{sec:model}, we describe the data samples and the model that we used for analyses. Section \ref{sec:result} presents the direct results from our modeling, namely the fraction of stars with \kepler-like planets, \fkep, and the average planet multiplicity, \npbar, as well as their dependencies on stellar effective temperature and thus stellar mass. In Section \ref{sec:discussion}, we compare our results with previous studies and discuss their implications. Finally, we summarize the paper in Section \ref{sec:summary}.

\section{Data Samples}{\label{sec:datasample}}

\subsection{The Stars} 
\label{sub:the_stars}
We select stars based on Table 1 of \citet{Berger_2018_ApJ_866_99B}, which revised the stellar properties of more than 170,000 \kepler\ targets using the \gaia\ data. 
In our study, we focus on stars that are flagged as main sequence in \citet{Berger_2018_ApJ_866_99B}, and thus exclude stars that are flagged as sub-giants or red giants or binary candidates based on \gaia\ radii. 
We only consider stars with effective temperatures between 3000 and 7500 K as stars outside this interval are very few.
Following \citet{Narang_2018_AJ_156_221N}, we make the sample cleaner by removing a few percent of outlier stars with a \kepler\ observation duty cycle less than 60\% or a data span shorter than 2 yr.
Applying above conditions, we obtain 101,159 stars in our star sample.
The sample size after each selection step is listed in Table \ref{table:select_data}.

\subsection{The Tranets} 
\label{sub:the_tranets}
We select transiting planet \citep[hereafter tranet;][]{Tremaine_2012_AJ_143_94T} candidates from the \kepler\ Data Release 25 \citep[DR25;][]{Thompson.2018ApJS..235...38T} catalog. 
There are 8054 \kepler\ Objects of Interest (KOIs) in DR25.
We remove KOIs that are flagged as false positives and only consider KOIs with host stars in our above star sample. We remove KOIs with orbital period $>$400 days, because \kepler\ detection efficiency rapidly drops at larger periods (Figure \ref{fig:fig_com_frac}).
We also remove those with abnormally large radii ($>$ 20 $R_{\earth}$).
In addition, following \citet{Mulders.2018AJ....156...24M}, we only consider highly reliable planet candidates with disposition scores $\geqslant 0.9$. 
Finally, we obtain 2509 planet candidates around 1889 stars in our tranet sample.
The sample size after each selection step is listed in Table \ref{table:select_data}.

\startlongtable
\begin{deluxetable}{l|cc}
\tablewidth{0pt}
\tablecaption{Data selection
\label{table:select_data} }
\tablehead{
\colhead{ } &\colhead{Star} &\colhead{Planet}}
\startdata 
	\kepler\ DR25           &199,991\tablenotemark{a}   &8054\\
	Not false positive                  &...            &4034\\
	Match with \gaia\ data              &177,911        &3642\\
	Main sequence                       &117,130        &3012\\
	Data span $\geqslant$ 2 yr       &107,054        &2983\\
	Duty cycle $\geqslant$ 0.6          &103,910        &2912\\
	No binary                           &101,159        &2892\\
	Period $\leqslant$ 400 days         &...            &2875\\
	$0<R_p \leqslant 20R_{\oplus}$      &...            &2825\\
	Disposition score $\geqslant$ 0.9   &...            &2509\\
\enddata
\tablenotetext{a}{We ignore 47 objects with KepIDs over 100,000,000, since not all of them are stars.}
\end{deluxetable}

\subsection{Tranet Distributions} 
\label{sub:tranet_distributions}
\begin{figure}[!thbp]
\includegraphics[width=.473\textwidth]{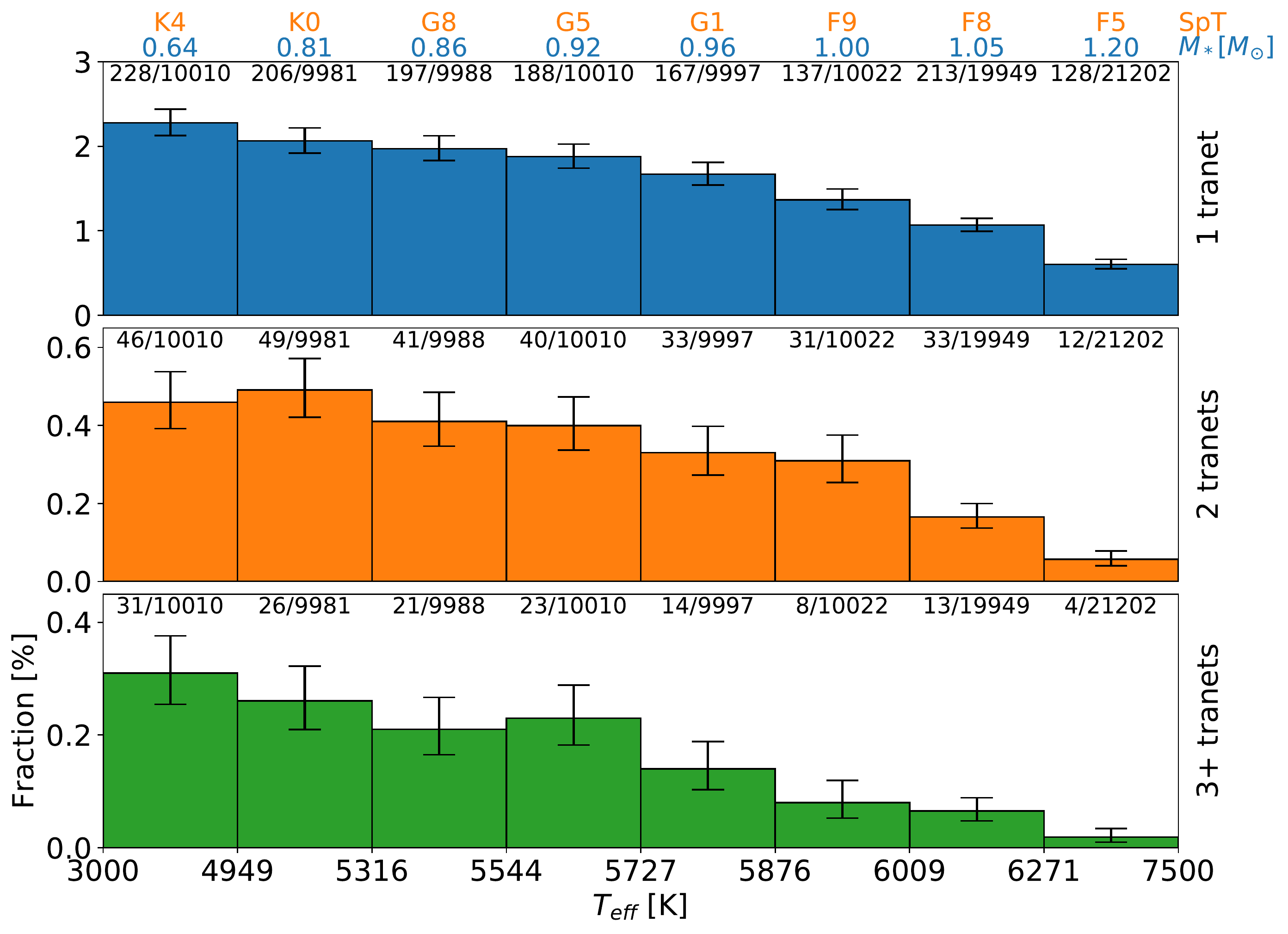}
\caption{Tranet fraction (the number of tranet systems divided by the number of stars as printed in each bin) as a function of stellar effective temperature. Systems with one, two, and three or more tranets are plotted in the top, middle, and bottom panels, respectively. The error bars assume the Poisson distribution in the counting uncertainties. At the top we also print the median stellar mass and the corresponding spectral type using the spectral--temperature relationship as in \citet{Pecaut_2013_ApJS_208_9P}.
\label{fig:fig_p_all_s}}
\end{figure}

\begin{figure}[!htp]
\includegraphics[width=.473\textwidth]{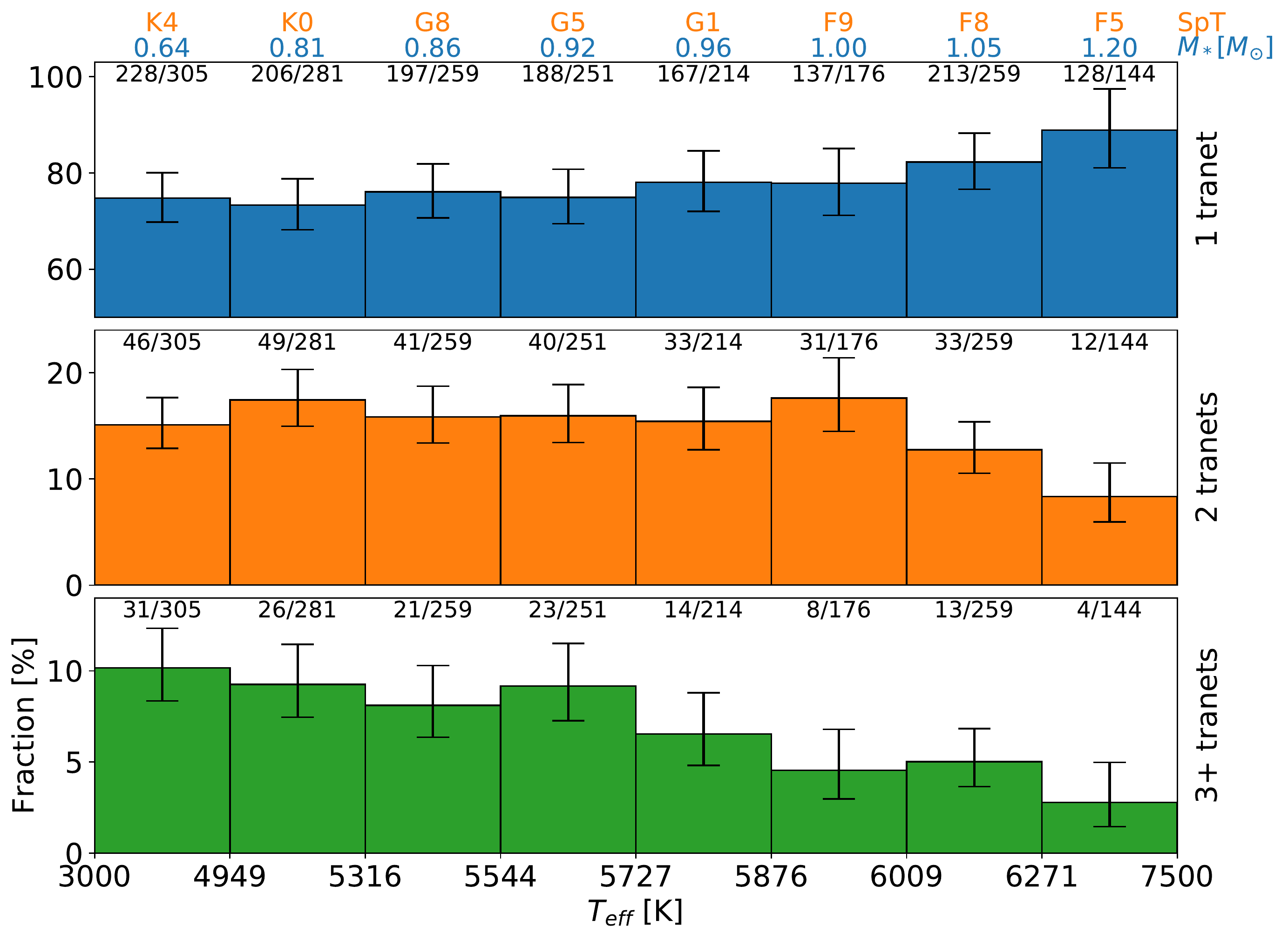}
\caption{Similar to Figure \ref{fig:fig_p_all_s}, but the $y$-axis is the contribution fraction (the number of tranet systems for a given multiplicity divided by the number of total tranet systems as printed in each bin).
\label{fig:fig_p_all_p}}
\end{figure}

\begin{figure}[!htp]
\includegraphics[width=.473\textwidth]{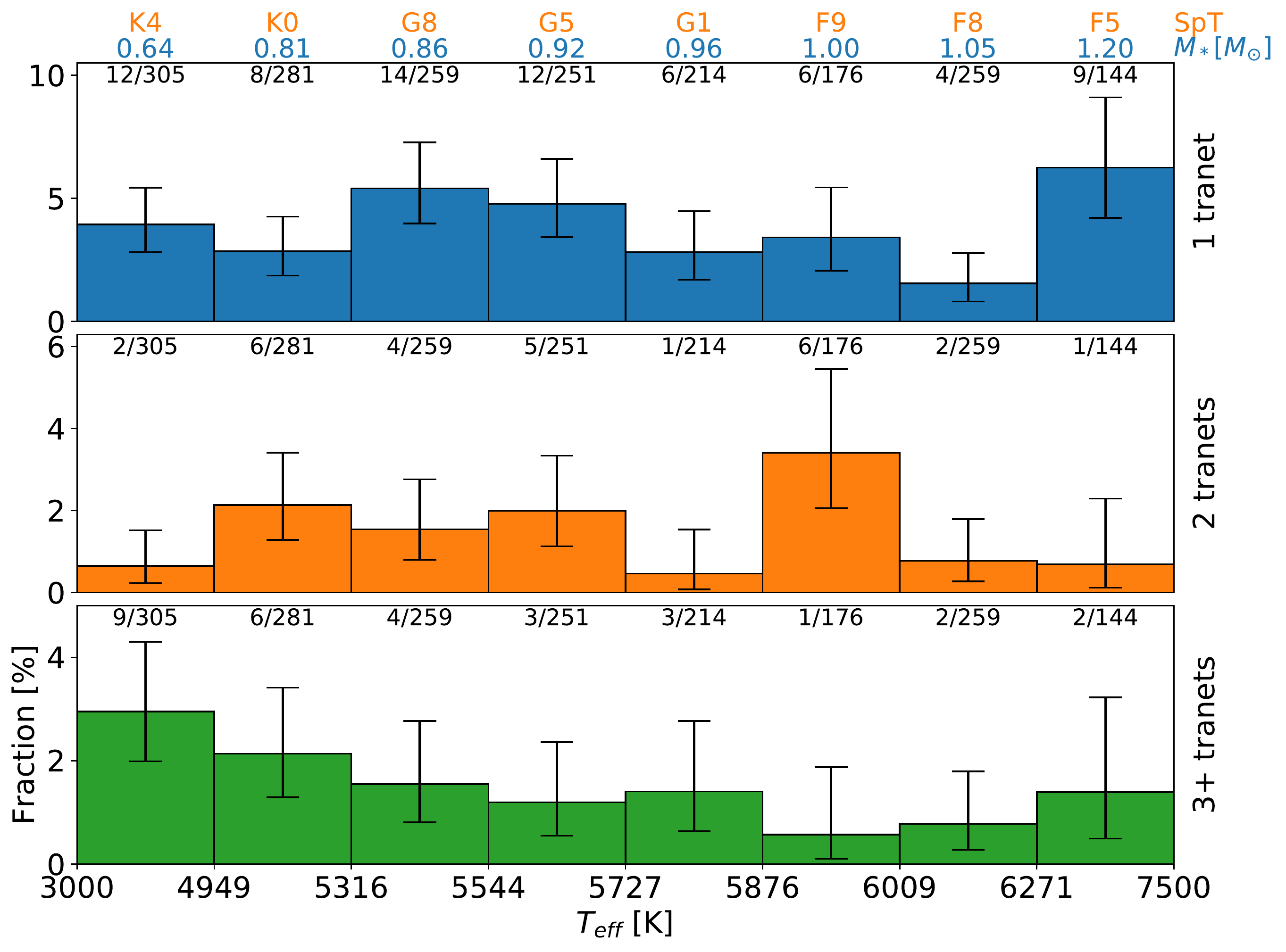}
\caption{Similar to Figure \ref{fig:fig_p_all_s}, but the $y$-axis is the TTV contribution fraction (the number of tranet systems with at least one tranet showing TTV signal for a given multiplicity divided by the number of total tranet systems as printed in each bin).
\label{fig:fig_ttv_all_p}}
\end{figure}

Since our star sample size is $\sim$100,000, we initially divided the star sample into 10 bins with approximately equal sizes ($\sim$10,000 stars in each bin) according to stellar effective temperature.
Nevertheless, the tranets in the last four bins are so few that we merge them into two bins.
We will show later that this shortage of tranets in the high temperature end is a combined effect of lower detection efficiency (Appendix \ref{sec:fdod_efficiency}) and lower intrinsic planet occurrence rate (Section \ref{sub:fkep} and \ref{sub:npbar}).
Figure \ref{fig:fig_p_all_s} shows the numbers of systems with one, two, and three or more tranets ($N_1$, $N_2$, $N_{3+}$) divided by the number of stars in each bin, respectively. 
We dub ($N_1$, $N_2$, $N_{3+}$) the tranet multiplicity function.
As can be seen, all the three panels reveal a common trend, namely, the tranet fraction decreases, by a factor $\sim$ 4, 8, and 16, respectively,  with increasing stellar effective temperature and thus stellar mass.
In Figure \ref{fig:fig_p_all_p}, we plot the contribution fractions of different tranet multiplicities in these eight bins. As can be seen, the contributions from higher (lower) tranet multiplicities generally decrease (increase) with increasing stellar effective temperature.
In Figure \ref{fig:fig_ttv_all_p}, we plot the transit-timing variation (TTV) fraction, i.e., the number of systems with at least one tranet showing TTV signals identified by \citet{Holczer_2016_ApJS_225_9H} divided by the number of tranet systems ($M_1$, $M_2$, $M_{3+}$)/($N_1+N_2+N_{3+}$) as a function of effective temperature. Here, we dub ($M_1$, $M_2$, $M_{3+}$) the TTV multiplicity function. 
In the following sections, we will build a model (Section \ref{sec:model}) to fit the above observed tranet distributions, which allow us to constrain the intrinsic occurrence and architecture of planet systems (Section \ref{sec:result} and \ref{sec:discussion}).

\section{Model}{\label{sec:model}}

\subsection{Overall Procedure} 
\label{sub:overall_procedure}
Our model is based on the framework of \citet{Zhu_2018_ApJ_860_101Z}, but with many modifications and new ingredients.
With the model, we generate tranet systems with model-expected tranet multiplicity function ($\bar{N}_1, \bar{N}_2, \bar{N}_{3+}$) and TTV multiplicity function ($\bar{M}_1, \bar{M}_2, \bar{M}_{3+}$). 
Since we have already divided the sample into different temperature bins, we bin the tranets into three groups instead of six as in \citet{Zhu_2018_ApJ_860_101Z} to avoid small number statistic. 
The simulated multiplicity and TTV functions are compared to the observed ones ($N_1, N_2, N_{3+}$ and $M_1, M_2, M_{3+}$) and the likelihood is computed as
\begin{equation}
	\mathcal{L} = \prod\limits_{k=1}^{3+} \frac{\bar{N}_k^{N_k} \exp (-\bar{N}_k)}{N_k !} \times
		\prod\limits_{k=1}^{3+} \frac{\bar{M}_k^{M_k} \exp (-\bar{M}_k)}{M_k !}.
\end{equation}

Then we apply \texttt{emcee} \citep{Foreman-Mackey.2013PASP..125..306F}, a \texttt{python} package using a Markov Chain Monte Carlo (MCMC) algorithm as the optimization method to constrain the free parameters, i.e., the fraction of stars with \kepler-like planets (\fkep; Section \ref{ssub:intrinsic_multiplicities}), the average planet multiplicity (\npbar; Section \ref{ssub:intrinsic_multiplicities}), and the inclination slope index ($\alpha$; Section \ref{ssub:orbital_inclinations}) of the model. We repeat this procedure for every temperature bin and therefore obtain \fkep, \npbar, and $\alpha$ as functions of stellar temperature. In the following, we describe the details of the model. 

\subsection{Individual Ingredients} 
\label{sub:individual_ingredients}

\subsubsection{Assuming intrinsic multiplicities} 
\label{ssub:intrinsic_multiplicities}

In \citet{Zhu_2018_ApJ_860_101Z}, the intrinsic multiplicities are modeled as six free parameters, i.e., $f_k$, the fraction of stars with $k$ \kepler-like planets, where $1\leqslant k \leqslant 6$. As found by \citet{Zhu_2018_ApJ_860_101Z}, individual $f_k$ were only loosely constrained, nevertheless, \fkep\ and \npbar\ could still be well constrained. For this reason, here in this work, we only consider \fkep\ and \npbar\ as the two free parameters in our model. 
We randomly select \fkep*$N_{\text{star}}$ stars, where $ N_{\text{star}}$ is the number of stars in a given temperature bin. For each selected star, we assign $k$ planets where $k$ is generated from a Poisson distribution with mean of \npbar\ as in \citet{Fang_2012_ApJ_761_92F}. Unlike \citet{Zhu_2018_ApJ_860_101Z}, here $k$ is cut off at 10, i.e., $1 \leqslant k \leqslant 10$. It is worth noting that our results are not sensitive to the choice of $k$ distribution. We performed some tests and found even if no distribution was assumed, i.e., $f_1$, $f_2$, …$f_6$ were all treated as free parameters, we still obtained similar constraint on \fkep\ and \npbar, after a much longer MCMC run.


\subsubsection{Assigning transit parameters and planet radii}
\label{ssub:epsilon_rp}
For each generated planet, we assign it a transit parameter (star radius divided by the semi-major axis of planet orbit: $\epsilon = R_*/a_p$) and a radius ($R_p$), which are drawn from the debiased distributions of observed  $\epsilon$ and $R_p$.
The \kepler\ transit survey generally involves three bias processes: transit geometry bias, detection efficiency bias, and vetting efficiency bias.
For the geometric debias, we give each tranet a weight ($1/f_{\text{tra}}$) that is the inverse of the transit probability, where $f_{\text{tra}}\sim\epsilon$. 
Here, we ignore the minor impact of orbital eccentricity since the majority of \kepler\ planets have small eccentricity ($<$0.1) as found by \citet{Xie_2016_PNAS_11311431X}.
For the detection debias, we give each tranet a weight ($1/f_{\text{S/N}}$), which is the inverse of the pipeline detection efficiency (see Figure \ref{fig:fig_com_frac} in Appendix \ref{sec:fdod_efficiency}) calculated by using \texttt{KeplerPORTs} \footnote{\url{https://github.com/nasa/KeplerPORTs}} \citep{Burke.2017ksci.rept...17B} with the detection metrics from the website of the exoplanet archive\footnote{\url{https://exoplanetarchive.ipac.caltech.edu/docs/Kepler_completeness_reliability.html}}. 
For the vetting debias, we give each tranet a weight ($1/f_{\text{vet}}$), which is the inverse of the KOI vetting efficiency derived by using the fitting result (their Equation 17) of \citet{Mulders.2018AJ....156...24M}. 
Therefore, each observed tranet has a total weight of $1/f_{\text{com}}$, where $f_{\text{com}}=f_{\text{tra}} \cdot f_{\text{S/N}} \cdot f_{\text{vet}}$ is the survey completeness, combining all the above three biases.

The debiased $\epsilon$ and $R_p$ distributions are treated as the approximation of intrinsic distributions from which we generate planets in our model.
As we will see in Figure \ref{fig:fig_5_compare} (Appendix \ref{sec:com_os}), our model generally reproduces the observed $\epsilon$ and $R_p$ distributions.

\subsubsection{Adjusting period ratios and radius ratios}
\label{ssub:pr_rr}
Above planet system generating processes assume the planets are randomly paired. 
In order to better match the observed period ratio \citep[pr;][]{Fabrycky.2014ApJ...790..146F, Brakensiek.2016ApJ...821...47B} and radius ratio (rr) distributions \citep{Ciardi.2013ApJ...763...41C, Weiss.2018AJ....155...48W},
we further adjust the orbit ratios and rrs of the generated planet systems. 
Before the adjustment, we first debias the observed pr and rr distributions.
For each observed adjacent tranet pair, we use the \texttt{CORBITS} algorithms \citep{Brakensiek.2016ApJ...821...47B} to calculate the probability of detecting the outer tranet given that the inner one is detected. 
The inverse of the probability is adopted as the weight of the tranet pair.

After obtaining the debiased distributions, we then use them to adjust the prs and rrs of the generated planet systems in our model.
Specifically, we first randomly select a planet in a given system. 
Next, from the debiased distributions, we draw prs and rrs and use them to adjust the periods and radii of the neighbouring planets. 
Then, such adjustments spread to other neighbouring planets until all planets go through.
As we will see in Figure \ref{fig:fig_5_compare} (Appendix \ref{sec:com_os}), our model generally reproduces the observed pr and rr distributions.
The results of switching off the adjustment are also discussed in Section \ref{sub:compare_to_zhu} and Appendix \ref{sec:different_bins}.

\subsubsection{Checking Orbital Stability}
\label{ssub:stability_criterion}
After we get $\epsilon$ for each planet, the corresponding orbital period is
\begin{equation}
	P=\left( \frac{R_{\odot}}{\text{au}} \right)^{3/2} \epsilon^{-3/2} \left( \frac{\rho_*}{\rho_{\odot}} \right)^{-1/2} \text{year}.
\end{equation}
For multiple-planet systems, there is another restriction that planets should not be too close to each other and become dynamically unstable. 
We adopt the \citet{Deck_2013_ApJ_774_129D} criterion, i.e., the pr of any planet pair should be larger than a critical value,
\begin{equation}
	\frac{P_{\text{out}}}{P_{\text{in}}} > 1+2.2q^{2/7},
\end{equation}
where $q$ stands for planet-star mass ratio.
We obtain stellar masses from the \kepler\ input catalog.
We calculate the planet masses based on their radii using the \texttt{python} package \texttt{Forecast} developed by \citet{Chen.2017ApJ...834...17C}. 
If any of the planet pairs do not satisfy stability criterion, we regenerate $\epsilon$ for all the planets in the system.
The stability check here mainly removed those unstable pairs with both large radii and small prs, though most unstable pairs would have already been removed if the pr adjustment (section \ref{ssub:pr_rr}) was taken.


\subsubsection{Assigning orbital inclinations to  generate transits} 
\label{ssub:orbital_inclinations}
For each system that passed the orbital stability check, we assign their planets, $I_p$, the orbital inclination relative to the observer. 
Following \citet{Zhu_2018_ApJ_860_101Z}, in practice,  we calculate

\begin{equation}
\cos I_p = \cos I \cos i \ -\ \sin I \sin i \cos \phi,
\end{equation}
where $I$ is the inclination of the system invariable plane, $i$ the planet inclination with respect to this invariable planet, and $\phi$ the phase angle. The distribution of $I$ is isotropic (i.e., $\cos I$ is uniform for $0^\circ <I<180^\circ$ ) and $\phi$ is random between 0$^\circ$ and 360$^\circ$. For single-planet systems, $i=0^\circ$ and $I_p=I$. For multiple-planet systems, following \citet{Zhu_2018_ApJ_860_101Z}, $i$ is modeled as a Fisher distribution,
\begin{equation}
P\left(i|\kappa_k\right)=\frac{\kappa_k \sin i}{2 \sinh \kappa_k}e^{\kappa_k \cos i}.
\end{equation}
The $\kappa_k$ parameter is related to the inclination dispersion as
\begin{equation}
\sigma_{i,k}^2 = \left<\sin^2 i\right> = \frac{2}{\kappa_k}\left(\coth \kappa_k - \frac{1}{\kappa_k}\right).
\end{equation}
Here, also following \citet{Zhu_2018_ApJ_860_101Z}, the inclination dispersion is a power-law function of the planet multiplicity, $k$, 
\begin{equation}\label{eq:inclination}
\sigma_{i,k}\equiv \sqrt{\left<\sin^2i\right>}=\sigma_{i,5}\left(\frac{k}{5}\right)^{\alpha}.
\end{equation}

By fitting the observed transit duration ratio, \citet{Zhu_2018_ApJ_860_101Z} found that the 1$\sigma$ confidence interval of $\sigma_{i,5}$ is between 0$\fdg$65 and 0$\fdg$96. 
In this paper, we adopt their result and draw $\sigma_{i,5}$ from a normal distribution with mean of 0$\fdg$8 and standard deviation of 0$\fdg$15. 
And $\alpha$ is treated as a free parameter with a uniform prior distribution between -4 and 0, which will be further constrained during the MCMC fitting process. 
The lower boundary of $\alpha$ is set as $-4$ because the inclination dispersion, $\sigma_{i,k}$, by its definition in Equation \ref{eq:inclination} has a maximum value of $\sqrt{2/3}$. 

The orbital inclination relative to the observer, $I_p$, together with the transit parameter, $\epsilon$, will finally determine whether a planet transits or not. 
Here a transit is defined as the impact parameter less than 1, i.e., $|\cos \left(I_p\right)/\epsilon|<1$. 
As in \citet{Zhu_2018_ApJ_860_101Z}, we ignore the minor impact of the planet size.


\subsubsection{Applying detection and vetting efficiencies} 
\label{ssub:transit_detection_criterion}

Not every transit contributes to observation depending on the transit detection efficiency ($f_{\text{S/N}}$) and KOI vetting efficiency ($f_{\text{vet}}$), which are calculated as in Section \ref{ssub:epsilon_rp} (see also in the Appendix \ref{sec:fdod_efficiency} for more discussions). 
For each transit generated from Section \ref{ssub:orbital_inclinations}, we draw a random number, $f_{\text{ran}}$, from a uniform distribution between 0 and 1. If $f_{\text{ran}}<f_{\text{S/N}} \cdot f_{\text{vet}}$, then this transit can be detected, and it finally contributes to the simulated tranet multiplicity function ($\bar{N}_1, \bar{N}_2, \bar{N}_{3+}$).


\subsubsection{Applying TTV detection criteria} 
\label{ssub:ttv_detection_criteria}
In order to obtain the simulated TTV multiplicity function ($\bar{M}_1, \bar{M}_2, \bar{M}_{3+}$), we apply TTV detection criteria to select TTV systems from the simulated tranet systems. The criteria are the same as in \citet{Zhu_2018_ApJ_860_101Z} and they are summarized as follows.
\begin{enumerate}
\item TTV signals associated with the first order of mean motion resonances ($J$:$J$-1 = 2:1, 3:2, 4:3, and 5:4) are considered.
\item The orbital periods of TTV planets should be less than 200 days.
\item The super period of the planet pair, $P_{\text{sup}}$, is in the range of 100$\leqslant P_{\text{sup}} \leqslant 3000$ days, where $P_{\text{sup}}$ is
\begin{equation}
	P_{\text{sup}} \equiv \frac{P_{\text{in}}P_{\text{out}}}{|J P_{\text{in}}-(J-1)P_{\text{out}}|}.
\end{equation}
\item TTV amplitude indicator $P/ \Delta > 1300$ days, where $\Delta$ represents the fractional separation to period commensurability \citep{Lithwick.2012ApJ...761..122L},
\begin{equation}
	\Delta \equiv \left|  \frac{P_{\text{out}}}{P_{\text{in}}}\frac{J-1}{J} -1 \right|.
\end{equation}
\end{enumerate}

\begin{figure}[!htp]
\includegraphics[width=.473\textwidth]{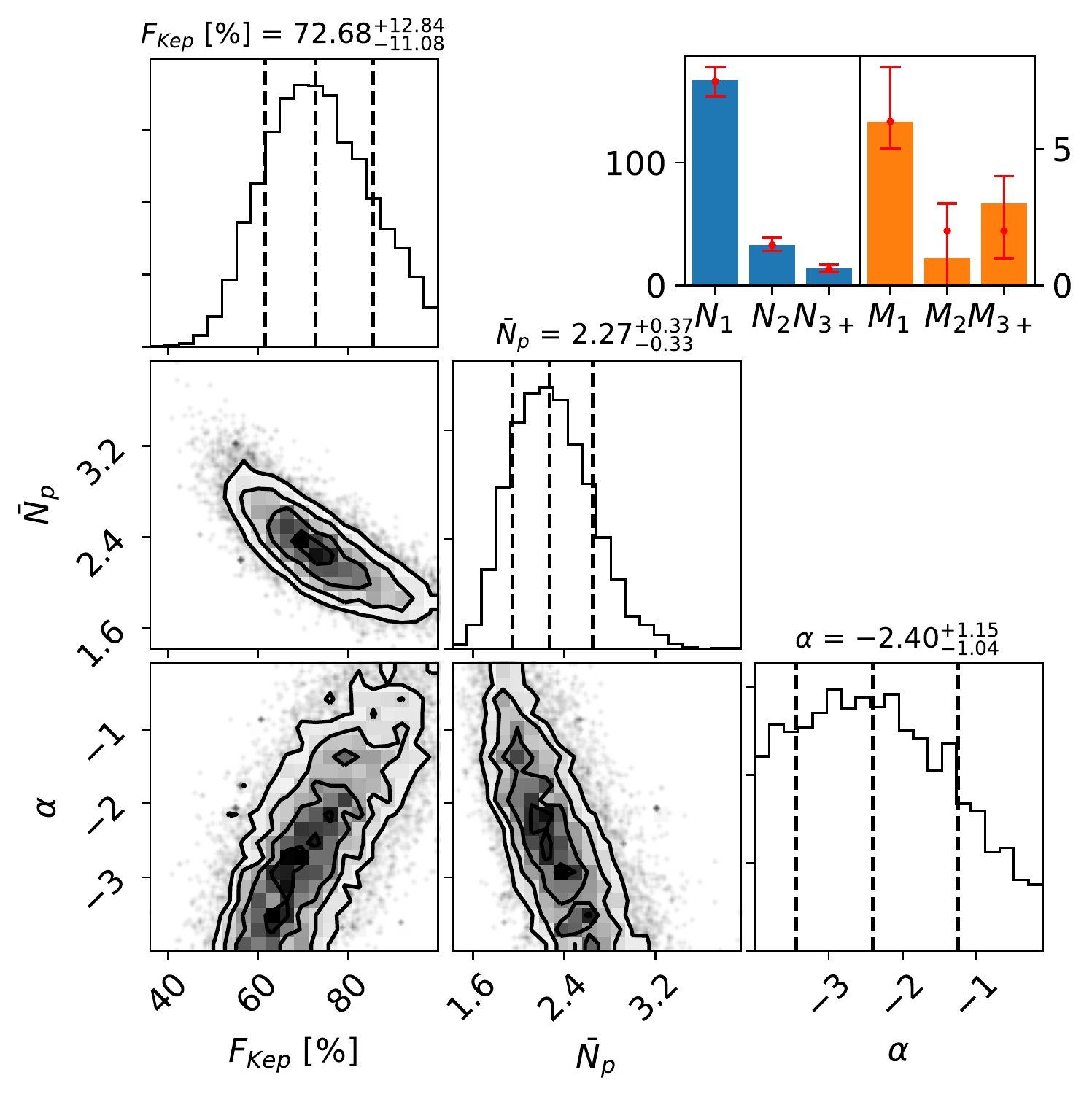}
\caption{Lower left: posterior distribution of \fkep, \npbar, and $\alpha$ from the fifth bin with \teff= 5727-5876 K (i.e., the MCMC corner plots). The figure is plotted using \texttt{corner.py} \citep{Foreman-Mackey.2016JOSS....1...24F}.
The dotted lines mark $50\pm34.1$ percentile positions, with the corresponding values labelled on the top.
Upper right: blue and orange bars show the observed multiplicity and TTV function, and the red circles and error bars show show $50 \pm 34.1$ percentiles of the posterior distributions of the fitting results.
\label{fig:fig_g1}}
\end{figure}

\section{RESULT}{\label{sec:result}}
\subsection{MCMC fit in a Solar-like \teff\ bin}{\label{sub:mcmc_solor}}
As mentioned in Section \ref{sub:overall_procedure}, by fitting the simulated tranet multiplicity ($\bar{N}_1, \bar{N}_2, \bar{N}_{3+}$) and TTV multiplicity function ($\bar{M}_1, \bar{M}_2, \bar{M}_{3+}$) with the observed tranet multiplicity ($N_1, N_2, N_{3+}$) and TTV multiplicity functions ($M_1, M_2, M_{3+}$), we are able to constrain our model parameters, i.e., the fraction of stars with \kepler-like planets (\fkep; Section \ref{ssub:intrinsic_multiplicities}), the average planet multiplicity (\npbar; Section \ref{ssub:intrinsic_multiplicities}), and the inclination slope index ($\alpha$; Section \ref{ssub:orbital_inclinations}), and obtain the function relations between them and stellar effective temperature (\teff).  
In Figure \ref{fig:fig_g1}, we plot the MCMC fitting results for the fifth \teff\ bin (G1 spectra on average). 
We choose it as an example simply because the Sun's temperature is in this bin. 
For the MCMC results of other \teff\ bins, please check Figure \ref{fig:fig_os_n} and \ref{fig:fig_os_m} in Appendix \ref{sec:com_os}.
As can be seen, \fkep\ and \npbar\ are well constrained to be Gaussian-like distributions, and they are somewhat anticorrelated. 
This anticorrelation is not unexpected, because it generally reflects the fact that increase/decrease in \fkep \ can somewhat compensate for the decrease/increase in \npbar\ to yield a given number of tranets.
Here, \fkep$\sim 73\%$ (\npbar$\sim 2.3$) is higher (lower) than that obtained by \citet{Zhu_2018_ApJ_860_101Z}.
This may be because the detection efficiency and vetting efficiency correction, as well as pr and rr adjustment, which were ignored in \citet{Zhu_2018_ApJ_860_101Z}, are all taken into account in this work (see Section \ref{sub:compare_to_zhu} for more discussion).
On the other hand, the inclination slope parameter, $\alpha$, is constrained to be toward the lower boundary -4, which is consistent with the results in \citet{Zhu_2018_ApJ_860_101Z}.
In the following subsections, we present the results of \fkep\,, \npbar\,, and $\alpha$ as functions of \teff.

\subsection{\fkep\ as a function of \teff}
\begin{figure}[!htp]
\includegraphics[width=.473\textwidth]{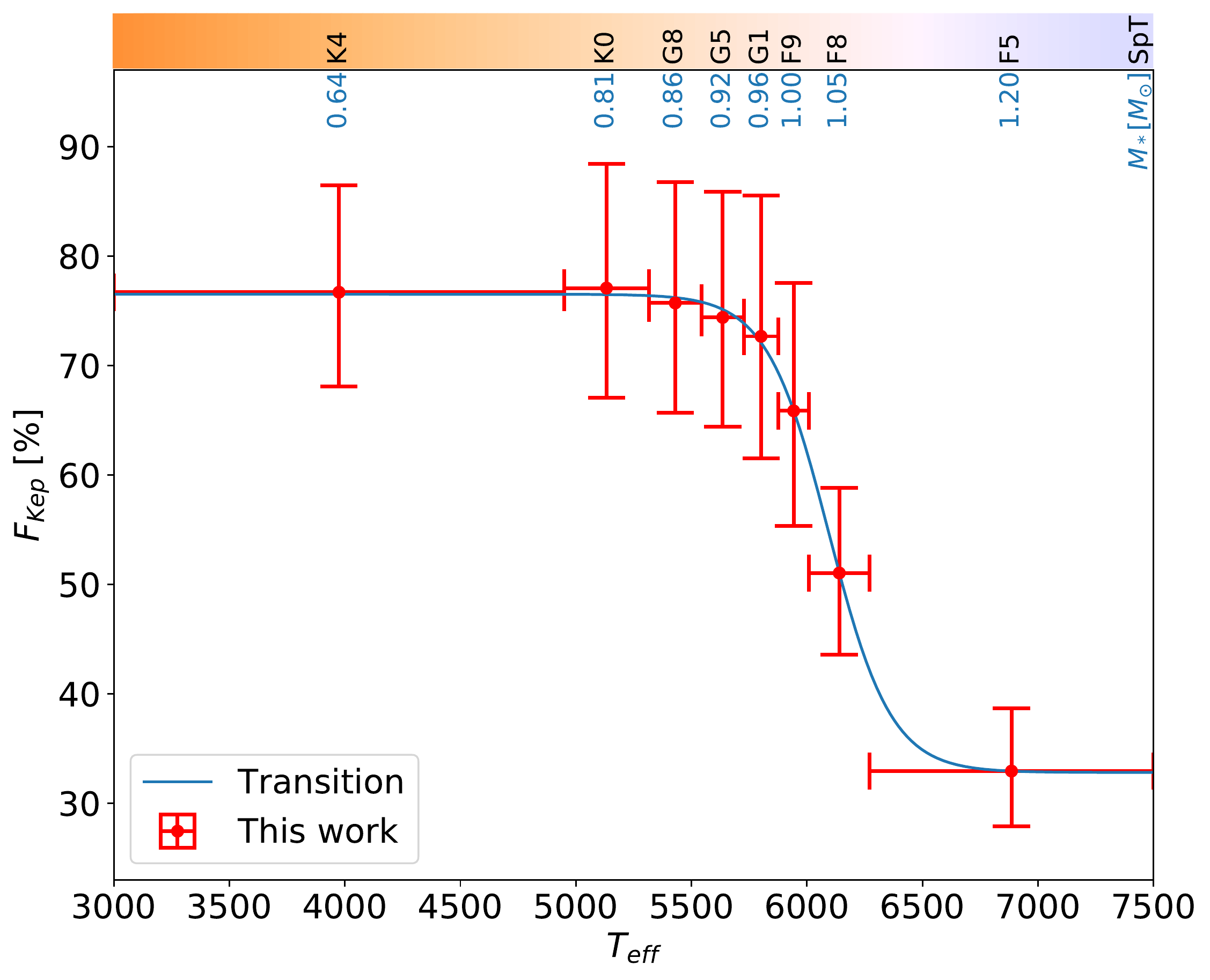}
\caption{Fraction of stars with \kepler\ planets (\fkep) as functions of stellar effective temperature. The circles and error bars indicate the $50\pm34.1$ percentiles from the posterior distributions after the MCMC fitting. The blue curve shows the best fit of the transition model, see Table \ref{table:fkep} for AIC scores of different models. At the top, we also print the median stellar mass and the corresponding spectral type using the spectral--temperature relationship as in \citet{Pecaut_2013_ApJS_208_9P}.
\label{fig:fig_fit_fkep}}
\end{figure}

{\label{sub:fkep}}
In Figure \ref{fig:fig_fit_fkep}, we plot the fraction of stars with \kepler-like planets, \fkep, as a function of stellar effective temperature, \teff\,.
As can be seen, \fkep\, decreases progressively and significantly with \teff\,. 
As \teff\ increases from 3000 K to 7500 K, \fkep\ decreases from $\sim$75\% to $\sim$35\%.
Such a striking trend is not unexpected because it is actually revealed by the observational fact shown in Figure \ref{fig:fig_p_all_s}, namely, the tranet fractions for all the subgroups strongly decline with increasing \teff\,.
In order to quantify the decline trend, we fit it with four different functions, namely a constant function ($f_1$), a linear function ($f_2$), a two-step function ($f_3$), and a transition function ($f_4$), whose formulae are given below.
\begin{eqnarray}
    f_1 &=& b \\
    f_2 &=& b\times T_{\text{eff}}+c \\
	f_3 &=& b \left( T_{\text{eff}} \leqslant T_0\right)\label{eq:2step}\,\, {\rm or} \,\, c \left( T_{\text{eff}} > T_0\right)\\
	f_4 &=& b + \frac{c}{1+\exp{\left(\dfrac{T_{\text{eff}}-T_0}{\Delta T}\right)}}\label{eq:4par}
\end{eqnarray}
The transition function has four free parameters, $b$, $c$, \teff, and $T_0$.
It looks complex, but it well describes where ($T_0$) and how quickly ($\Delta T$) the transition from $b$ to $b+c$ takes place.

For each function, we calculate the Akaike information criterion (AIC) score of the best fit, which is listed in Table \ref{table:fkep}. 
As can be seen, the transition function is most preferred with the lowest AIC score of 8.4. 
The formula of the best fit is 
\begin{eqnarray}
	F_{\text{Kep}}&=&0.30 +
	 \frac {0.43}{1+\exp{\left( \dfrac{T_{\text{eff}}-6061 }{161}\right)}},
\end{eqnarray}
which is overplotted in Figure \ref{fig:fig_fit_fkep}.
The linear and two-step functions give slightly larger AIC scores, 9.7 and 11.7 respectively, indicating that they are statistically comparable to the transition function.
For clarity, we only show the model with the lowest AIC score.
In contrast, the constant function gives a much larger AIC score, 41.9.
The AIC difference is so large ($\Delta \rm AIC>$30) that the constant function can be securely excluded.
This quantitatively demonstrates the strong decline trend of \fkep\ with \teff. 

\startlongtable
\begin{deluxetable}{ccccc}
\tablewidth{0pt}
\tablecaption{AIC scores of different fitting results for \fkep, \npbar\ and inclination slope index ($\alpha$). \label{table:fkep} }
\tablehead{
\colhead{Parameter}	&\colhead{Constant}	&\colhead{Two-step}	&\colhead{Linear}	&\colhead{Transition}}
\startdata 
\fkep 			&41.9			&9.7				&11.7			&8.4\\
\npbar			&11.6			&6.3				&6.7			&8.5\\
$\alpha$		&3.7			&5.7			    &5.3			&9.2
\enddata
\end{deluxetable}

\subsection{\npbar\ as a function of \teff}
\begin{figure}[!htp]
\includegraphics[width=.473\textwidth]{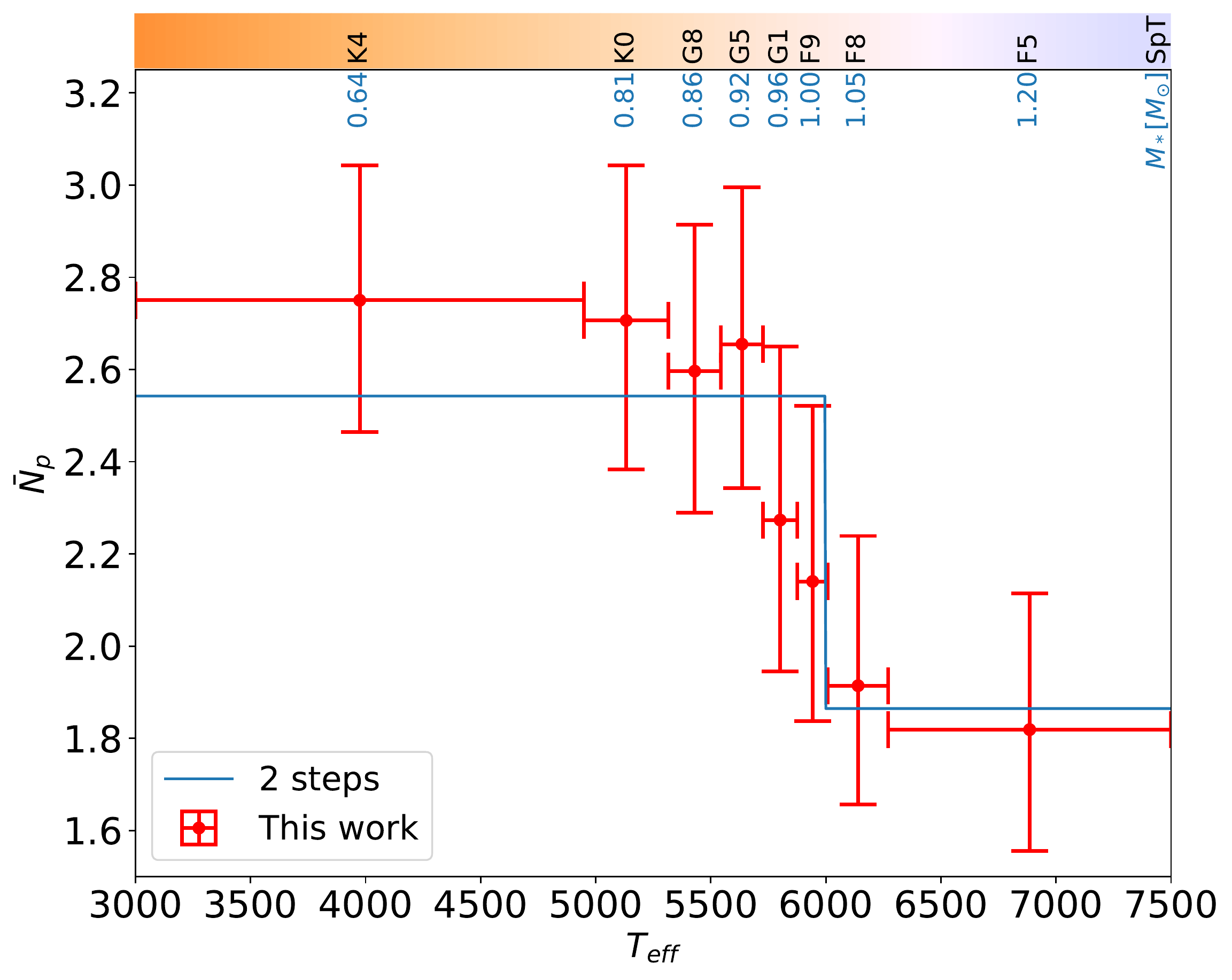}
\caption{Average planet multiplicity (\npbar) as function of stellar effective temperature. The circles and error bars indicate the $50\pm34.1$ percentiles from the posterior distributions after the MCMC fitting. The blue curve shows the best fit of the two-step model, see Table \ref{table:fkep} for the AIC scores of different models.
\label{fig:fig_fit_npbar}}
\end{figure}

{\label{sub:npbar}}
In Figure \ref{fig:fig_fit_npbar}, we plot the average planet multiplicity, \npbar, as a function of stellar effective temperature, \teff. 
As can be seen, \npbar\ also decreases, though not as significant as \fkep\ with increasing \teff.
Such a decline trend is also not unexpected because it is actually revealed by the observational fact shown in Figure \ref{fig:fig_p_all_p}, namely, the relative fraction of multiple tranets (e.g., systems with three or more tranets as shown in the bottom panel of Figure \ref{fig:fig_p_all_p}) decreases with increasing \teff.
As the error bars in Figure \ref{fig:fig_p_all_p} are relatively larger than those in Figure \ref{fig:fig_p_all_s}, one may expect that the significance of the \npbar\ decline trend is lower than that of the \fkep\ decline trend, which is revealed by Figure \ref{fig:fig_p_all_s}.
To quantify the the \npbar\ decline trend, we perform the same analysis as in Section \ref{sub:fkep} for the \fkep\ decline trend.
We find the most preferred function to fit the \npbar\ decline trend is a two-step function.
The best fit formula is 
\begin{eqnarray}
\bar{N}_p&=&2.6\,\left(T_{\text{eff}} \leqslant 6000\ K\right)\, {\rm or}\,  1.9\,\left( T_{\text{eff}} > 6000\ K\right),
\end{eqnarray}
which gives an AIC score of 6.3 (Table \ref{table:fkep}) and it is overplotted in Figure \ref{fig:fig_fit_npbar}.
The linear and transition function give slightly larger AIC scores of 6.7 and 8.5, respectively, indicating that they are statistically comparable to the two-step function.
The constant function gives the largest AIC score of 11.6. 
The AIC difference is $\Delta \rm AIC$=5.3 between the constant and two-step functions, indicating that the \npbar\ decline trend is tentative, much less significant than the \fkep\ decline trend whose $\Delta \rm AIC>$30.

\subsection{$\alpha$ as a function of \teff}
\begin{figure}[!htp]
\includegraphics[width=.473\textwidth]{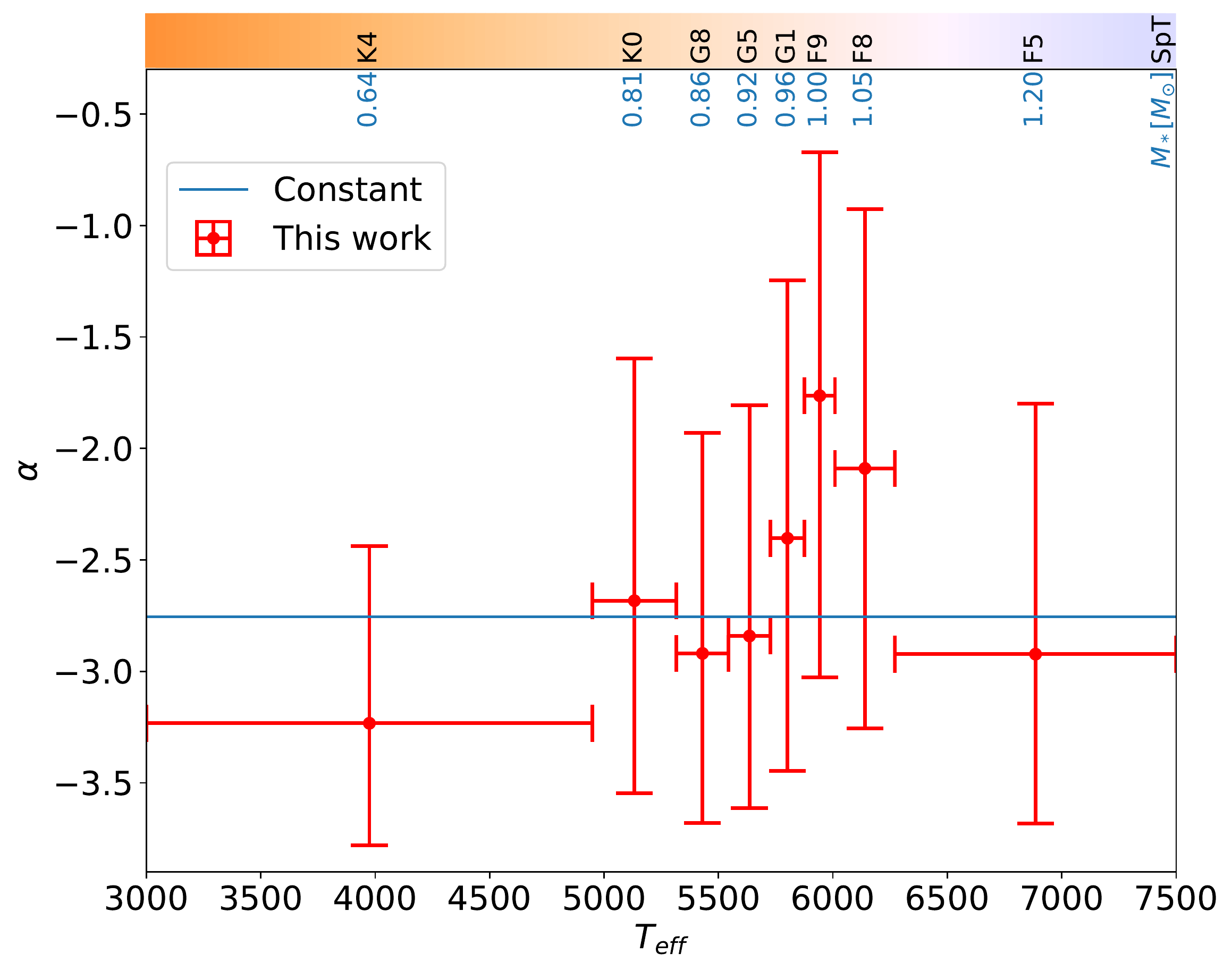}
\caption{Inclination slope index ($\alpha$) as function of stellar effective temperature. The circles and error bars indicate the $50\pm34.1$ percentiles from the posterior distributions after the MCMC fitting. The blue curve shows the best fit of the constant model, see Table \ref{table:fkep} for the AIC scores of different models.
\label{fig:fig_fit_alpha}}
\end{figure}

{\label{sub:alpha}}
In Figure \ref{fig:fig_fit_alpha}, we plot the inclination slope index ($\alpha$, defined in Equation \ref{eq:inclination} in Section \ref{ssub:orbital_inclinations}) as a function of stellar effective temperature \teff.
As can be seen, $\alpha$ has no obvious change trend with \teff\,.
And it is fit best with a constant function, namely $\alpha =-2.8$, which gives the lowest AIC score of 3.7 as shown in Table \ref{table:fkep}. 
Also note the error bars of $\alpha$ as shown in Figure \ref{fig:fig_fit_alpha}, which are relatively large. 
This is because $\alpha$ is mainly constrained by the TTV multiplicity function ($M_1, M_2, M_{3+}$), which is relatively uncertain due to the small TTV sample size in each \teff\ bin. 

\begin{figure}[!htp]
\includegraphics[width=.473\textwidth]{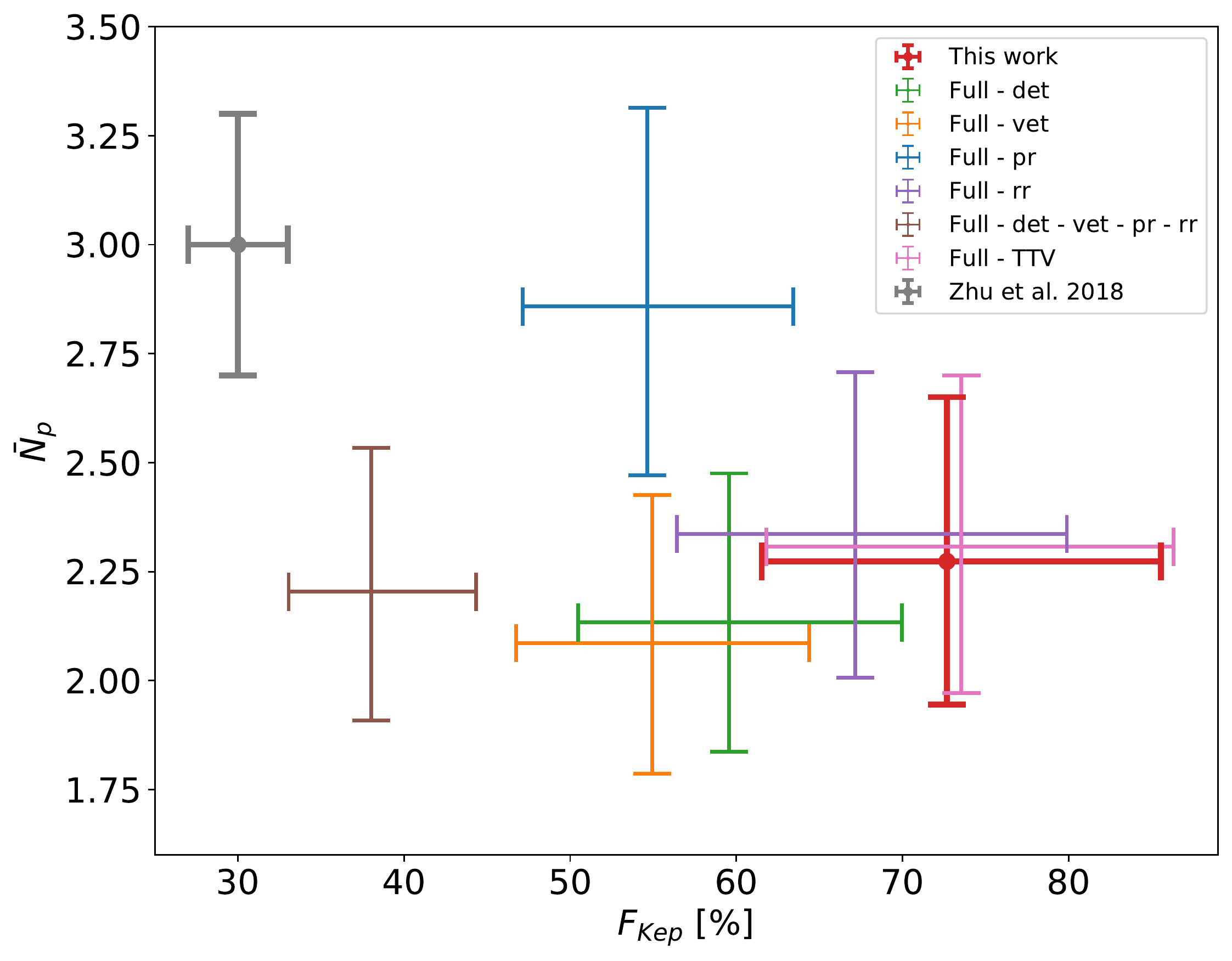}
\caption{Comparison on the \fkep--\npbar\ plane between the results of our work (the fifth bin with \teff$\sim5727-5876$ K) and that of \citet[][the grey symbol]{Zhu_2018_ApJ_860_101Z}. 
The red symbol is for the result of our full model with all the ingredients as mentioned in Section \ref{sub:individual_ingredients} being taken into account. 
Other symbols with different colors are the results of different modified models with some ingredients being removed with respect to the full model. The legend names "full-det", "full-vet", "full-pr", "full-rr", "full-pr-rr-det-vet", and "full-TTV" stand for models in which detection efficiency correction (det), vetting efficiency correction (vet), period ratio adjustment (pr), radius ratio adjustment (rr), detection and vetting efficiency corrections as well as period and radius ratio adjustments, and TTV multiplicity fitting (TTV) are removed with respect to the full model, respectively.
\label{fig:fig_bin5}}
\end{figure}

\section{Discussions} 
\label{sec:discussion}

\subsection{Comparison to \citet{Zhu_2018_ApJ_860_101Z}}
\label{sub:compare_to_zhu}

We compare our \fkep-\npbar\ results to that (the grey symbol) of \citet{Zhu_2018_ApJ_860_101Z} in Figure \ref{fig:fig_bin5}.
The red symbol is for the result of our full model, with all the model ingredients (Section \ref{sub:individual_ingredients}) being taken into account. 
As can be seen, for the average planet multiplicity, our result (\npbar\,$\sim2.3\pm0.4$) is a bit lower than theirs (\npbar\,$\sim3.0\pm0.3$).
However, for the fraction of stars with \kepler\ planets, our result (\fkep$\sim73\%\pm13\%$) is much larger than theirs (\fkep$\sim30\%\pm3\%$).
There are many reasons for the large difference in \fkep\ between the two studies.
First, the data sets are different. 
For the tranet sample, \citet{Zhu_2018_ApJ_860_101Z} used the \kepler\ Data Release 24, while we used the final Data Release 25. 
For the star sample, they select $\sim$ 30,000 solar-type stars based on a wide temperature range (\teff=4700-6500 K) and a cut on surface gravity (log $g>4$) given by LAMOST \citep{Luo.2015RAA....15.1095L}, while our result in Figure \ref{fig:fig_bin5} is for $\sim$ 10,000 solar-type stars selected from the HR diagram \citep{Berger_2018_ApJ_866_99B} with a narrower temperature range (\teff$\sim5700-5900$ K).
Second, the models are different. 
As mentioned in Section \ref{sec:model}, although our model is based on the framework of \citet{Zhu_2018_ApJ_860_101Z}, we added a number of ingredients that were not considered in their model, including rr and pr adjustments and detection (det) and vetting (vet) efficiency corrections.
As shown in Figure \ref{fig:fig_bin5}, removing any one of these model ingredients more or less changes \fkep.
As expected, ignoring the \kepler\ detection efficiency (Full-det) and KOI vetting efficiency (Full-vet) somewhat underestimate \fkep\ (the green and orange symbols).
Without pr adjustments (Full-pr), one also underestimates \fkep\ (the blue symbol). 
This is because random pairing would generate more planet pairs with large prs (e.g., pr$>$4), which increases their chances to be observed as singles.
To balance this effect and thus to fit the observed relative portion of single and multiple tranets, \npbar\ must increase somewhat, which then forces \fkep\ to decrease correspondingly (the \fkep-\npbar\ anticorrelation as shown in Figure \ref{fig:fig_g1}) to fit the absolute numbers of all tranets.
Without rr adjustment (Full-rr), it will cause similar effect, since rrs too high or too small (e.g., rr$>$5 or rr$<$0.2) also increase the chances of detecting multiple planets as singles (the purple symbol). 
If we remove all above ingredients to have a model (Full-det-vet-pr-rr) close to the one of \citet{Zhu_2018_ApJ_860_101Z}, then we get a \fkep$\sim38\%\pm6\%$ (the brown symbol), which is comparable to their result (\fkep$\sim30\%\pm3\%$). 
In addition, we find that whether or not including TTV (Full-TTV) in our model has little effect on \fkep\ and \npbar\ (the pink symbol). 
The effect of TTV, as also found by \citet{Zhu_2018_ApJ_860_101Z}, is mainly on constraining the $\alpha$ parameter. 

\subsection{Comparison to \citet{Howard_2012_ApJS_201_15H} and \citet{Mulders_2015_ApJ_798_112M}} 
\label{sub:number_of_planets_per_star}

\begin{figure}[!htp]
\includegraphics[width=.473\textwidth]{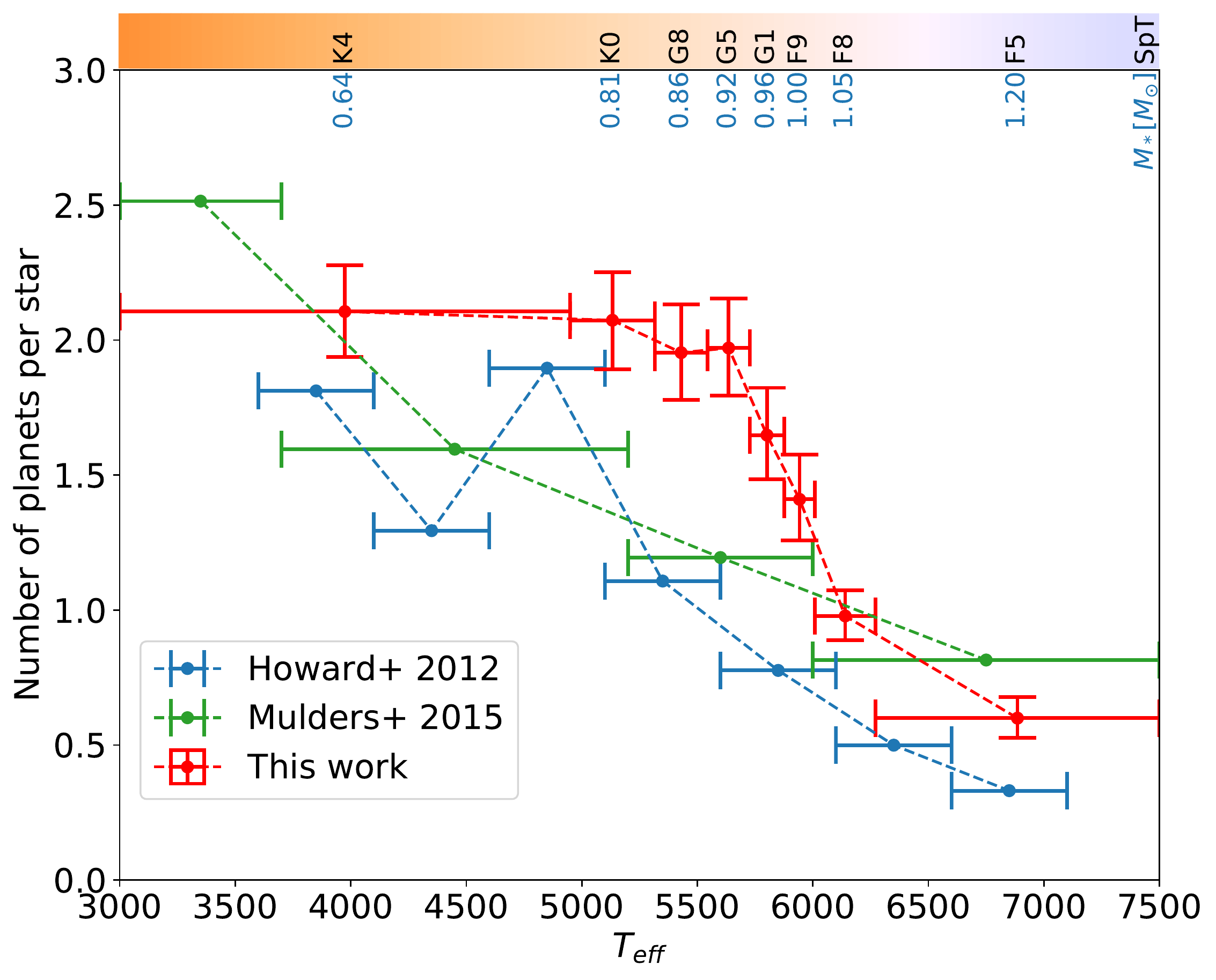}
\caption{Number of planets per star. We also compare our result with \citet[][blue line]{Howard_2012_ApJS_201_15H} and \citet[][green line]{Mulders_2015_ApJ_798_112M}. 
For fairness of comparison, we have extrapolated their results to the same planet ranges (i.e., planet radius $\sim 0.4-20 R_{\oplus}$ and period $<400$ days) as in our work. 
The extrapolations simply assume that the distributions of the planet occurrence rate keep the generally flat trend (as seen in \citet{Howard_2012_ApJS_201_15H} and \citet{Mulders_2015_ApJ_798_112M}) at a period $>50$ days and at a radius $<2R_{\oplus}$.
\label{fig:fig_numperstar}}
\end{figure}

We compare our results to those of \citet{Howard_2012_ApJS_201_15H} and \citet{Mulders_2015_ApJ_798_112M}, which studied the relation between stellar effective temperature (or mass) and planet occurrence rate but in terms of the average number of planets per star rather than the fraction of stars with planets (this work). 
The conversion between these two kinds of occurrence rates (Equation \ref{eq:fpnp}) becomes straightforward since the average planet multiplicity (the average number of planets per system with planets (\npbar), as well as the fraction of stars with planets (\fkep)), is already a direct product of our analysis. 
We convert our occurrence rates to the average number of planets per star using Equation \eqref{eq:fpnp} and compare them to those of \citet{Howard_2012_ApJS_201_15H} and \citet{Mulders_2015_ApJ_798_112M} in Figure \ref{fig:fig_numperstar}. 
As can be seen, our results are generally consistent with theirs, confirming the trend that the occurrence rate decreases with increasing stellar effective temperature. 
The average number of planets per star at the lower temperature end (2.1 for $<$4000 K) is a factor of 3.5 larger than that at the upper temperature end (0.60 for 7000 K). 
Nevertheless, we note there are differences in some specific details. 
On the one hand, the result of \citet{Mulders_2015_ApJ_798_112M} shows that the decrease in $\eta_p$ is likely to be fitted with a linear function of \teff\,.
On the other hand, the results of \citet{Howard_2012_ApJS_201_15H} and this work suggest that there seems to be a break point in the $\eta_p$ decreasing trend around \teff=5000 K, namely, $\eta_p$ decreases less (more) significantly for \teff\ lower (higher) than 5000 K. 
The reason for these subtle differences is not clear because different works used \kepler\ data of different release versions and adopted different statistical methods.
Recently, \citet{Garrett.2018PASP..130k4403G} used a different approach to model the occurrence rate ($\eta_p$) as a function of \teff\ whose results are also largely comparable to those shown in Figure \ref{fig:fig_numperstar}.

\subsection{Orbital Inclination and Obliquity} \label{sub:inclination_dis}

\begin{figure*}[!htp]
\centering
\includegraphics[width=\textwidth]{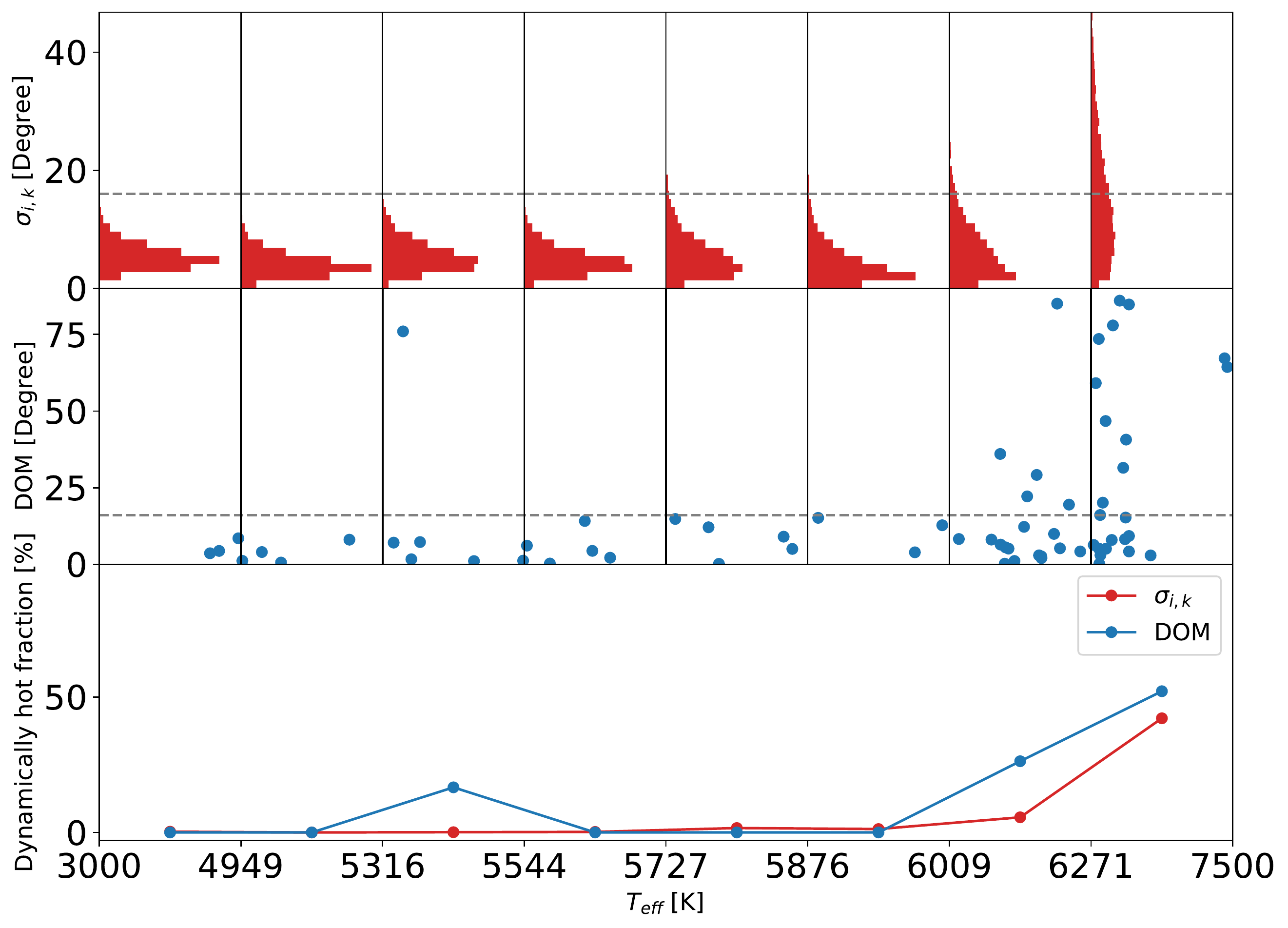}
\caption{Comparison between the inclination (top panel, in terms of $\sigma_{i,k}$, see Equation \ref{eq:inclination}) and the obliquity (middle panel, in terms of degree of misalignment, DOM, see the text in Section \ref{sub:inclination_dis}).  In the bottom panel, we plot the fractions of dynamically hot systems (defined as $\sigma_{i,k}$ or DOM greater than $16^{\circ}$, the horizontal dashed lines in the other two panels) as a function of \teff,. The boundary is chosen as it is  approximately the upper envelope  of the left part of the DOM distribution. We see that both $\sigma_{i,k}$ and DOM show a similar trend with \teff\,.   
\label{fig:fig_sl}}
\end{figure*}

One of the advantages of adopting the fraction of stars with planets as the planet occurrence rate is that it provides insight into the architecture of the underlying planetary systems. 
We can use Equation \eqref{eq:inclination} to derive the distribution of orbital inclination dispersion $\sigma_{i,k}$, since the power-law index, $\alpha$, and the number of planets in each system have already been obtained through the MCMC fitting. 
The result is plotted in upper panel of Figure \ref{fig:fig_sl}.
In the middle panel, we plot the degree of orbital misalignment (DOM) as a function of stellar effective temperature for observed systems, i.e., a modified version of the Figure 6 of \citet{Triaud.2018haex.bookE...2T}. 
For the sake of comparison to $\sigma_{i,k}$, DOM is defined as DOM = $\lambda$ if $\lambda<=90^\circ$ or DOM = 180$^\circ$--$\lambda$ if $\lambda>90^\circ$, where $\lambda$ is obliquity. 
As can be seen, for the six temperature bins on the left (\teff\,$<6000$ K), most $\sigma_{i,k}$ and DOM  are confined below $16^{\circ}$ (i.e., the horizontal dashed line).
In contrast, for the two temperature bins on the right (\teff\,$>6000$ K), there is a significant portion of dynamically hot systems with $\sigma_{i,k}$ or DOM above the horizontal dashed lines.
In the bottom panel, we plot the fraction of those dynamically hot ones (above the horizontal dashed lines) as a function \teff\,. 
Interestingly, both $\sigma_{i,k}$ and DOM show a similar trend.
The rise of $\sigma_{i,k}$ in the two higher \teff\, bins is expected because larger inclination dispersion reduces the observed transiting multiplicity, which naturally explains the falling of multiple tranets relative to single tranets as shown in Figure \ref{fig:fig_p_all_p}.
Nevertheless, the similarity between $\sigma_{i,k}$ and DOM as shown in Figure \ref{fig:fig_sl} is somewhat surprising. 
Below, we further discuss its implications.

The $\sigma_{i,k}$--temperature trend shown here (upper panel of Figure \ref{fig:fig_sl}) provides a complement to the known obliquity--temperature trend (middle panel of Figure \ref{fig:fig_sl}), and the similarity between them may suggest a common origin. 
The obliquity (DOM) trend is mainly from the observations of hot Jupiters with orbital periods $<$10 days \citep{Schlaufman.2010ApJ...719..602S, Winn.2010ApJ...718L.145W,Albrecht.2012ApJ...757...18A}.
One scenario for explaining the trend is that high obliquities are primordial at the beginning via some obliquity generation mechanisms, and cool stars are more capable of spin-orbital realigning than hot stars via tidal and/or magnetic effects \citep{Winn.2010ApJ...718L.145W, Dawson.2014ApJ...790L..31D, Spalding.2015ApJ...811...82S}. 
However, these theories suffer from problems in explaining the $\sigma_{i,k}$--temperature trend here. 
The latter is for \kepler\ planets, the bulk of which are super-Earths/sub-Neptunes residing in intrinsically multiple-planet systems with orbital periods $>$10 days, where tidal effects are negligible. 
Another possible scenario could be that the generation of obliquity or inclination differs for stars of different temperatures. 
Hotter stars are more likely to have giant planets \citep{Johnson_2010_PASP_122_905J,Ghezzi.2018ApJ...860..109G} and stellar companions \citep{Duchene.2013ARA&A..51..269D} whose dynamical interactions can excite obliquity \citep{Wu.2003ApJ...589..605W, Chatterjee.2008ApJ...686..580C, Wu.2011ApJ...735..109W, Batygin.2012Natur.491..418B} and simultaneously reduce planet multiplicity (which also explains Figure \ref{fig:fig_fit_npbar}). 
On the other hand, cooler stars have less giant planets and less stellar companions \citep{Duchene.2013ARA&A..51..269D}. 
Without the gravitational excitation from giant planets and companion stars, these systems are likely to stay in near coplanar orbits and low obliquities with high planet multiplicities. 
Future quantitative studies with detail modeling are needed to explain both the trends together.

If orbital inclination dispersion ($\sigma_{i,k}$) dominates obliquity, then the $\sigma_{i,k}$--temperature trend allows us to predict the obliquity distribution of \kepler\ planets.
Specifically, according to Figure \ref{fig:fig_sl}, we expect that planetary systems are generally well aligned with low obliquities $\sim 5^\circ$ for cool stars (\teff$<$6000 K) but mildly misaligned with moderate obliquities $\sim 10^\circ-20^\circ$ for hot stars \teff$>$6000 K). 
Our prediction of low obliquities for cool stars is consistent with the results of two recent studies, which found that the amplitudes of photometric variability \citep{Mazeh.2015ApJ...801....3M} and the projected rotation velocities \citep[$v\sin i$,][]{Winn.2017AJ....154..270W} of \kepler\ planet host stars are a factor $\sim 4/\pi$ higher than those of stars without detected transiting planet candidates, as one would expect if the planet hosts have low obliquities and the others are randomly oriented. However, for hot stars, there are discrepancies between the studies. On the one hand, \citet{Mazeh.2015ApJ...801....3M} found an opposite result for hot stars (\teff$>$6200 K); the amplitudes of photometric variabilities of \kepler\ planet hosts are a factor of $\sim 4/\pi$ lower than those of other stars. If the low amplitudes are caused by high obliquities, then the average obliquity of \kepler\ planet hosts would be even higher than $\arccos(\pi/4) \sim 38\fdg2$. On the other hand, \citet{Winn.2017AJ....154..270W} found that, similar to cool stars, the mean $v\sin i$ of hot \kepler\ planets hosts is also larger by a factor $\sim 4/\pi$ than that of other stars, and both cool and hot \kepler\ hosts have mean obliquity smaller than $\sim 20^\circ$. Our prediction of moderate obliquities ($\sim 10^\circ-20^\circ$) for hot stars is not contradictory to the result of \citet{Winn.2017AJ....154..270W}, though we further predict that the obliquities of hot stars are larger than those of cool stars by $10^\circ-20^\circ$. Such an obliquity difference ($<20^\circ$) only causes (2-6)\% variation in $v \sin i$, which could not be distinguished by \citet{Winn.2017AJ....154..270W}. Nevertheless, future surveys and missions, particularly the ongoing \textit{Transiting Exoplanet Survey Satellite} (TESS), are expected to detect many more planets around nearby bright stars, which are suitable for measuring obliquity in high precision and thus can test our prediction.

\section{Summary}\label{sec:summary} 
In this paper, we investigate the occurrence (in terms of fraction of stars with planets) and architecture of systems with \kepler\ planets (generally, radii $R_p \ga 0.4 R_{\oplus}$ and orbital period $P <$400 days) as a function of stellar effective temperature (or stellar mass equivalently for main-sequence stars). 
We find the following results.
\begin{enumerate}

\item The fraction of stars with \kepler-like planets decreases from \fkep$\sim$75\% for late-type stars with lower temperature (\teff$<$5000 K) to \fkep$\sim$35\% for early-type stars with higher temperature (\teff$>$6500 K; Figure \ref{fig:fig_fit_fkep}).

\item The average intrinsic planet multiplicity shows a tentative decreasing trend from \npbar$\sim$2.8 for late-type stars with lower temperature (\teff$<$5000 K) to \npbar$\sim$1.8 for early-type stars with higher temperature (\teff$>$6500 K; Figure \ref{fig:fig_fit_npbar}).

\item Orbital inclination dispersion is a steep falling function of planet multiplicity (i.e., generally systems with fewer planets are dynamically hotter, Equation \ref{eq:inclination}), with the power-law index, $\alpha\sim<-2$, which is nearly independent of stellar effective temperature (Figure \ref{fig:fig_fit_alpha}).
\end{enumerate}

We also discuss the effects of various ingredients of our models (i.e., detection efficiency and vetting efficiency corrections, pr and rr adjustments, and etc.).
Although these effects could cause significant differences in the absolute values of \fkep\, and \npbar\, (Figure \ref{fig:fig_bin5}), their relative trends with \teff\, remain generally the same (Figures \ref{fig:fig_fkep}-\ref{fig:fig_alpha}). 
The occurrence-temperature trend studied here is mainly for smaller planets, e.g., super-Earths and/or sub-Neptunes, and thus it is complementary to previous studies of the trend for giant planets \citep{Johnson_2010_PASP_122_905J, Ghezzi.2018ApJ...860..109G}.

Finally, we have revealed an inclination-temperature trend, which is similar and complementary to the well-known obliquity-temperature trend (Figure \ref{fig:fig_sl}). 
If the two trends share a common origin, then it may suggest that the generation of obliquity or inclination differs for stars of different temperatures.
Based on these trends, we predict that hot stars ($\ga$6000 K) with \kepler-like planets generally have slightly larger (by $10^\circ-20^\circ$) obliquities than those of cool ones. 
Future obliquity measurements in high precision will test this prediction.   

\label{sec:acknowledgments}

\acknowledgments
We thank W. Zhu for helpful comments and suggestions. This work is supported by the National Key R\&D Program of China (No. 2019YFA0405101) and the National Natural Science Foundation of China (NSFC) (grant No. 11933001, 11661161014). J.-W.X. also acknowledges the support from the National Youth Talent Support Program and the Distinguish Youth Foundation of Jiangsu Scientific Committee (BK20190005) 

\appendix
\section{Detection Efficiencies of Different \teff\, Bins} 
\label{sec:fdod_efficiency}
\begin{figure*}[ht!]
\centering
\includegraphics[width=.9\textwidth]{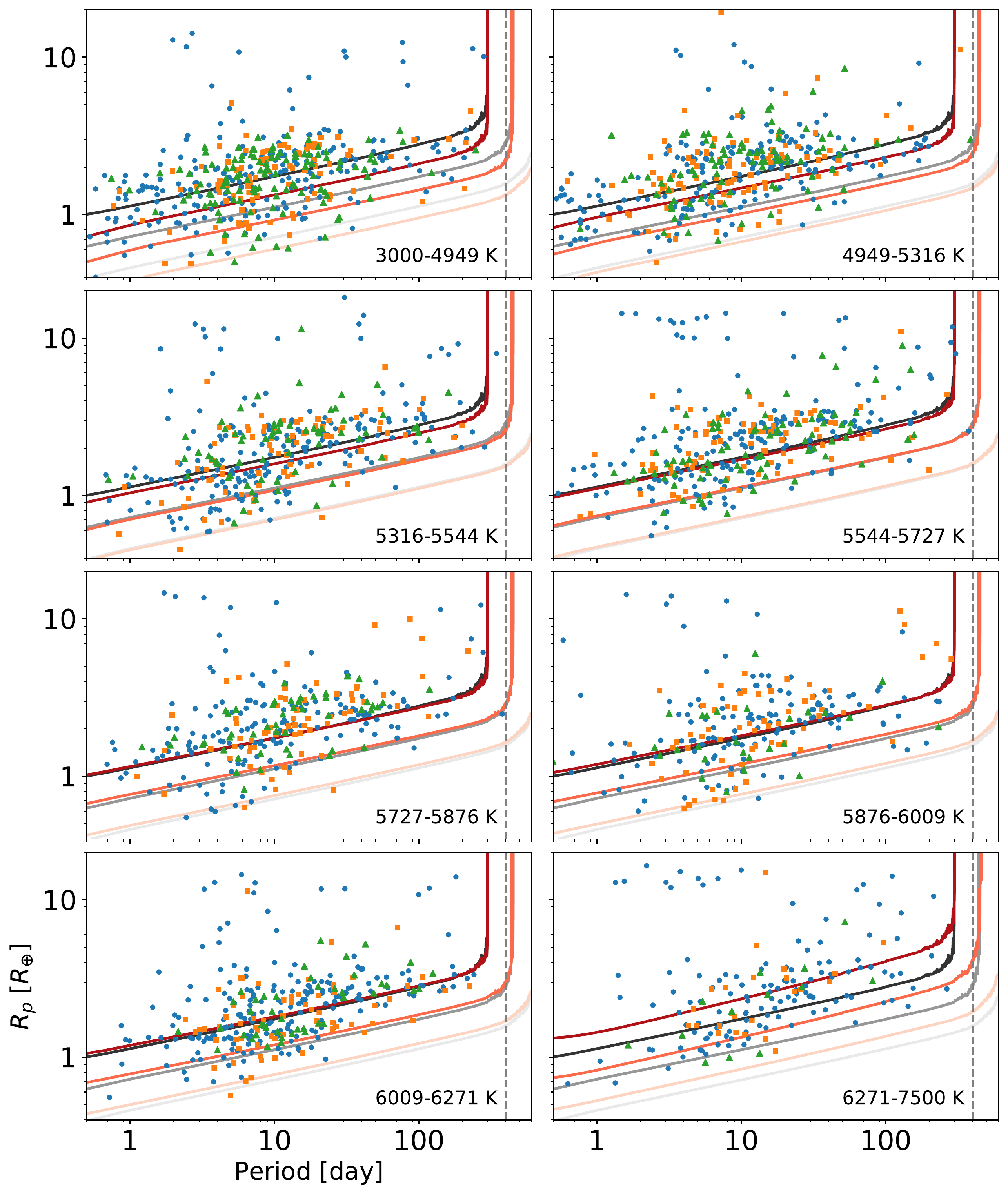}
\caption{Orbital periods vs. radii for the eight nominal bins of tranets in our sample. Systems with one, two, and three or more tranets are shown with blue dots, orange squares, and green triangles, respectively. In each panel, we plot their average detection efficiencies (10\%, 50\%, and 90\%) for stars in each bin (red lines). For easy of comparison, we also plot the same mean detection efficiencies for all stars in the whole sample (black lines). As can be seen, except for the first and the last bins, the detection efficiencies are close to each other. The vertical dashed line indicates the period cutoff at 400 days. 
\label{fig:fig_com_frac}}
\end{figure*}

\begin{figure*}[ht!]
\centering
\includegraphics[width=.8\textwidth]{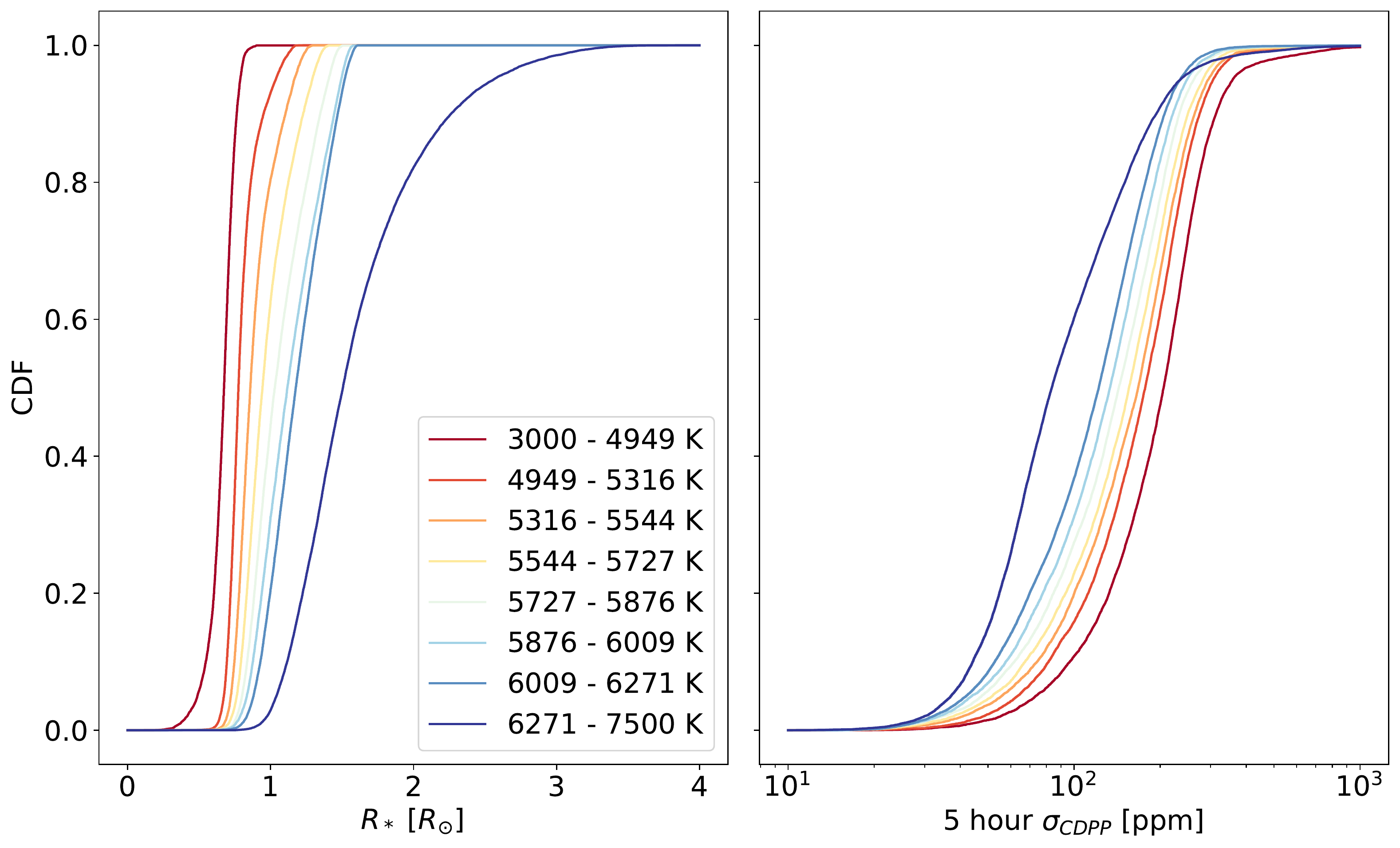}
\caption{Cumulative distributions of stellar radii (left) and the 5 hr $\sigma_{\text{CDPP}}$ (right) for stars in different bins. 
As can be seen, hotter (cooler) stars are larger (smaller) in sizes but with lower (higher) $\sigma_{\text{CDPP}}$, i.e., better (worse) photometry precision.
\label{fig:fig_rs_cdpp05p0}}
\end{figure*}

We calculate the transit detection efficiency by using the KeplerPORTs \citep{Burke.2017ksci.rept...17B} and the detection metrics available from the NASA exoplanet archive (https://exoplanetarchive.ipac.caltech.edu/docs/).
Figure \ref{fig:fig_com_frac} shows the 10\%, 50\%, and 90\% average detection efficiency contours as well as the tranet distributions in the period--radius diagram for the eight stellar temperature bins. 
As can be seen, the 10\% detection efficiency contours generally match the bottom envelopes of the tranet distributions, which are reasonable indications of their detection limits. 
We also note that the detection efficiency contours of different bins, except for the first and last bins, are close to each other.
Naively, one expects that planets are more easily detected around cooler stars, which are smaller in size (left panel of Figure \ref{fig:fig_rs_cdpp05p0}) and thus resulting in larger transit depths. 
Nevertheless, on the other hand, cooler stars are usually fainter with poorer photometric precision (larger $\sigma_{\text{CDPP}}$, right panel of Figure \ref{fig:fig_rs_cdpp05p0}) and thus reducing the transit S/N ratio. 
These two effects compensate each other, resulting in close detection efficiencies for stars with a large range of stellar effective temperature. 
Detection efficiencies in the first \teff\ bin are higher, and in the last \teff\ bin are lower than those in other bins.
This is consistent with the result of \citet{Christiansen.2015ApJ...810...95C}, which found that stars with \teff\,$<$4000 K and \teff\,$>$7000 K have different detection efficiencies than those with \teff$=4000-7000$ K.

\section{More Detailed Results}
\label{sec:com_os}
\begin{figure*}
\gridline{\fig{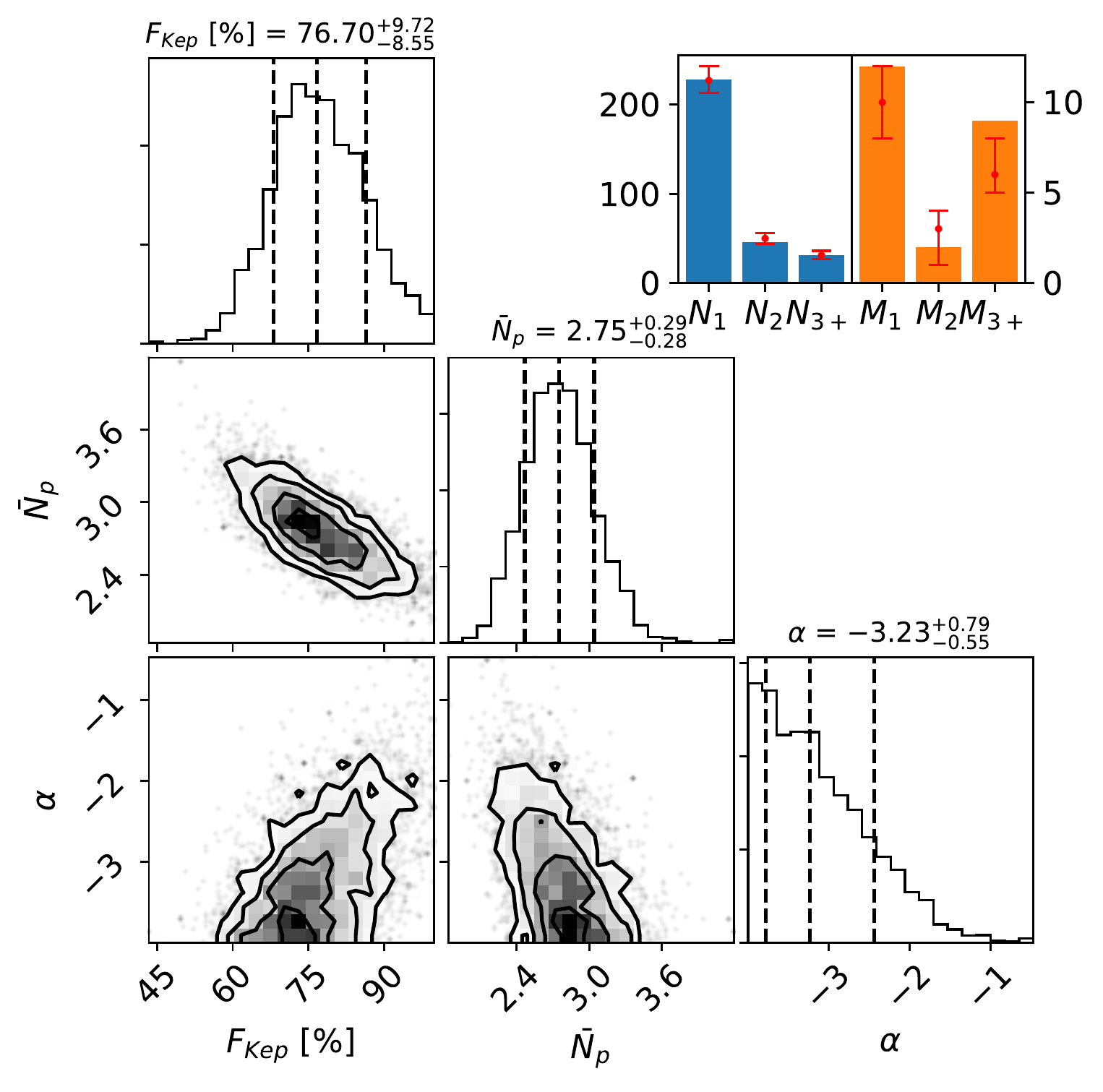}{0.5\textwidth}{Bin 1 (3000-4949 K)}
          \fig{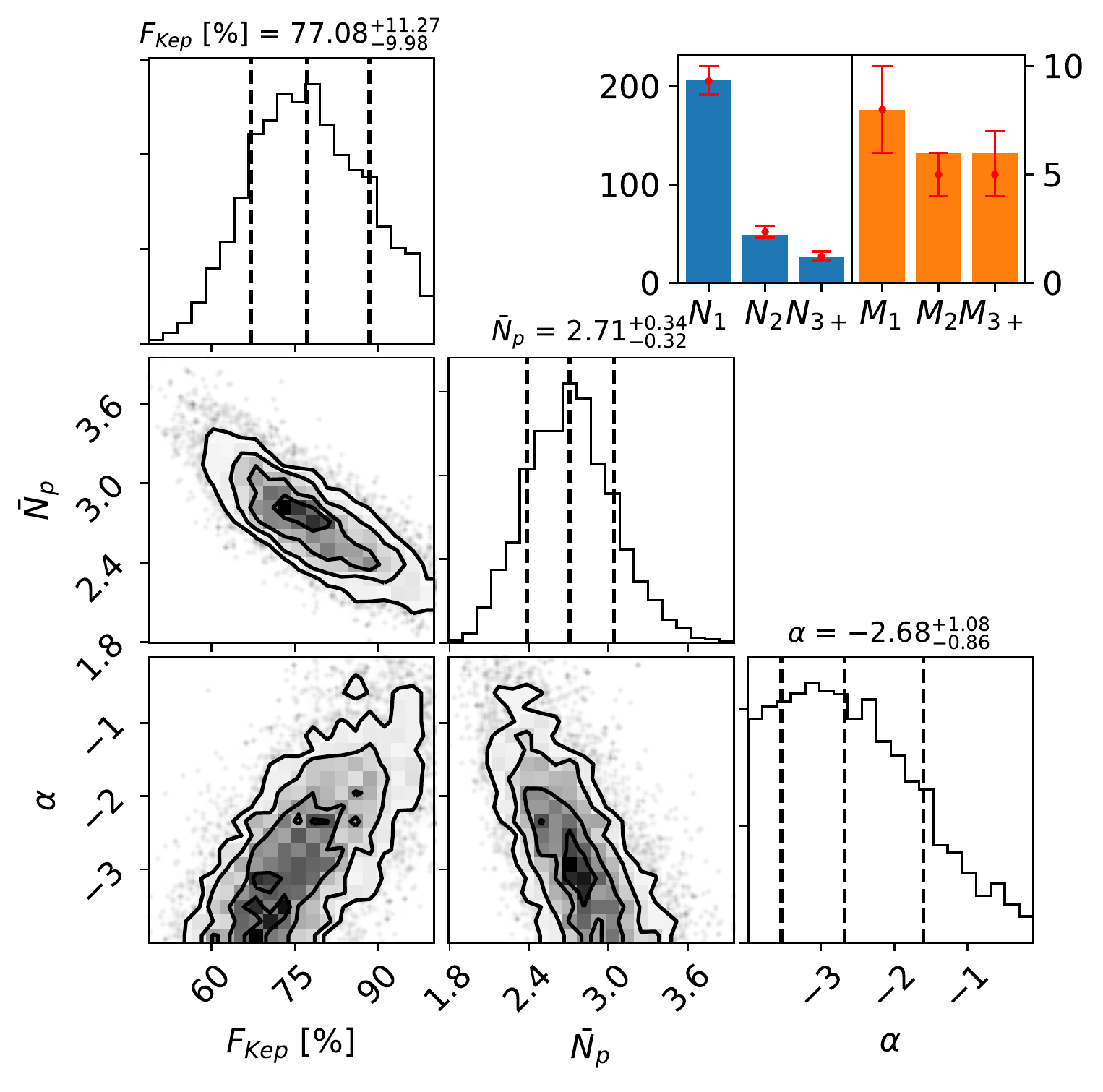}{0.5\textwidth}{Bin 2 (4949-5319 K)}
          }
\gridline{\fig{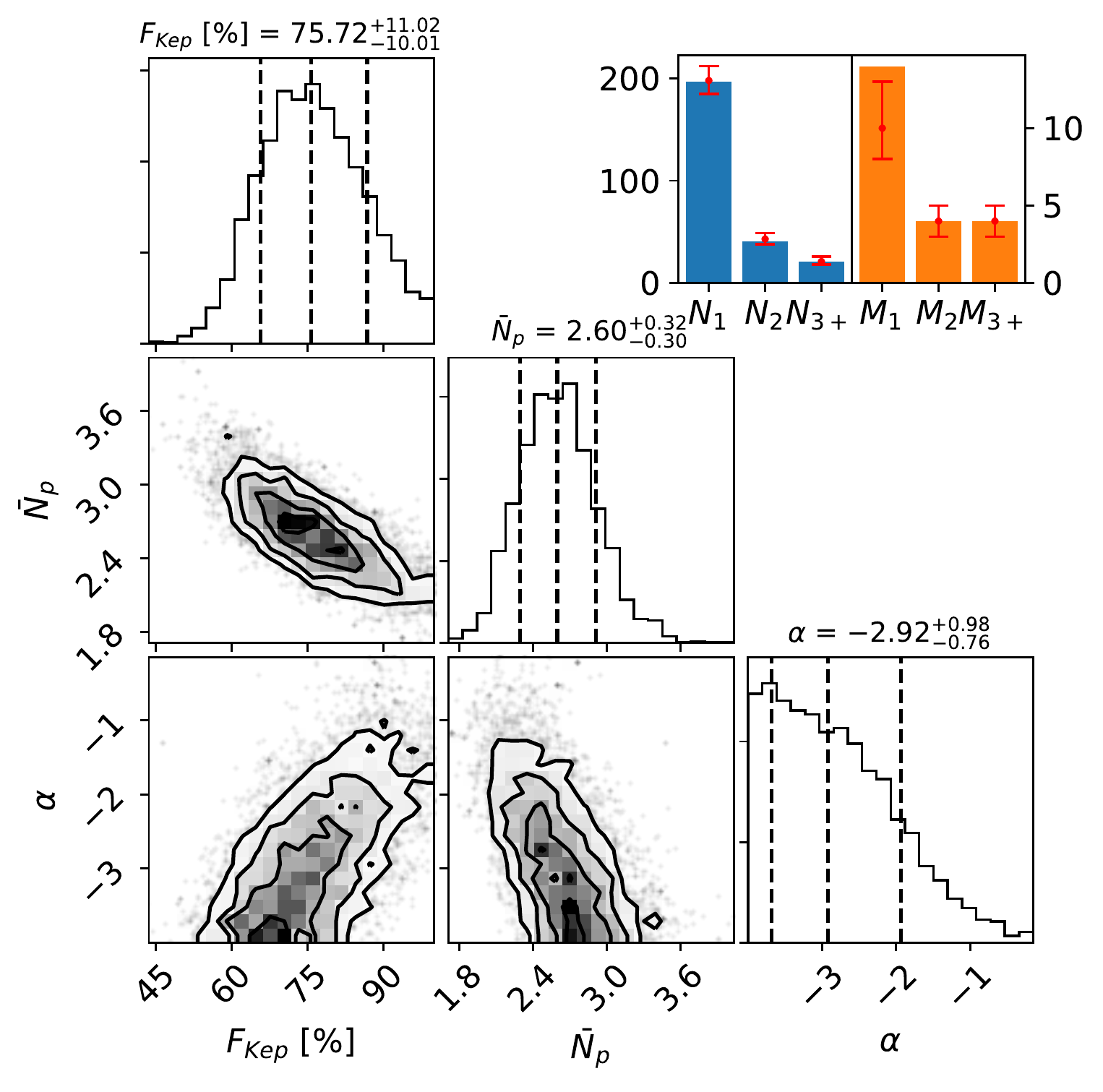}{0.5\textwidth}{Bin 3 (5319-5544 K)}
          \fig{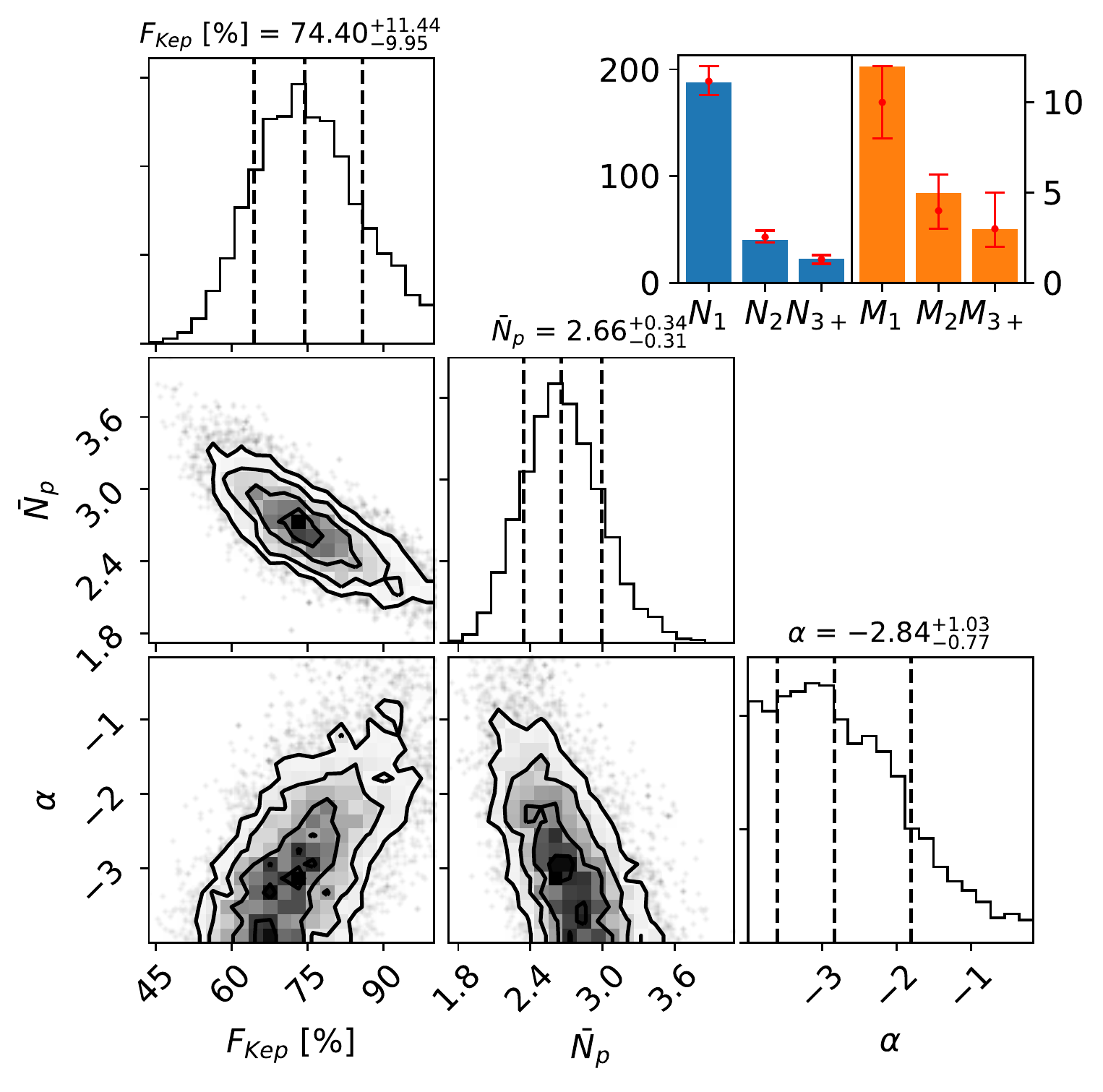}{0.5\textwidth}{Bin 4 (5544-5727 K)}
          }
\caption{Similar to Figure \ref{fig:fig_g1}, here are the MCMC results of the \teff\, bins 1-4.  \label{fig:fig_os_n}}
\end{figure*}

\begin{figure*}
\gridline{\fig{fig_g1_04.pdf}{0.5\textwidth}{Bin 5 (5727-5876 K)}
          \fig{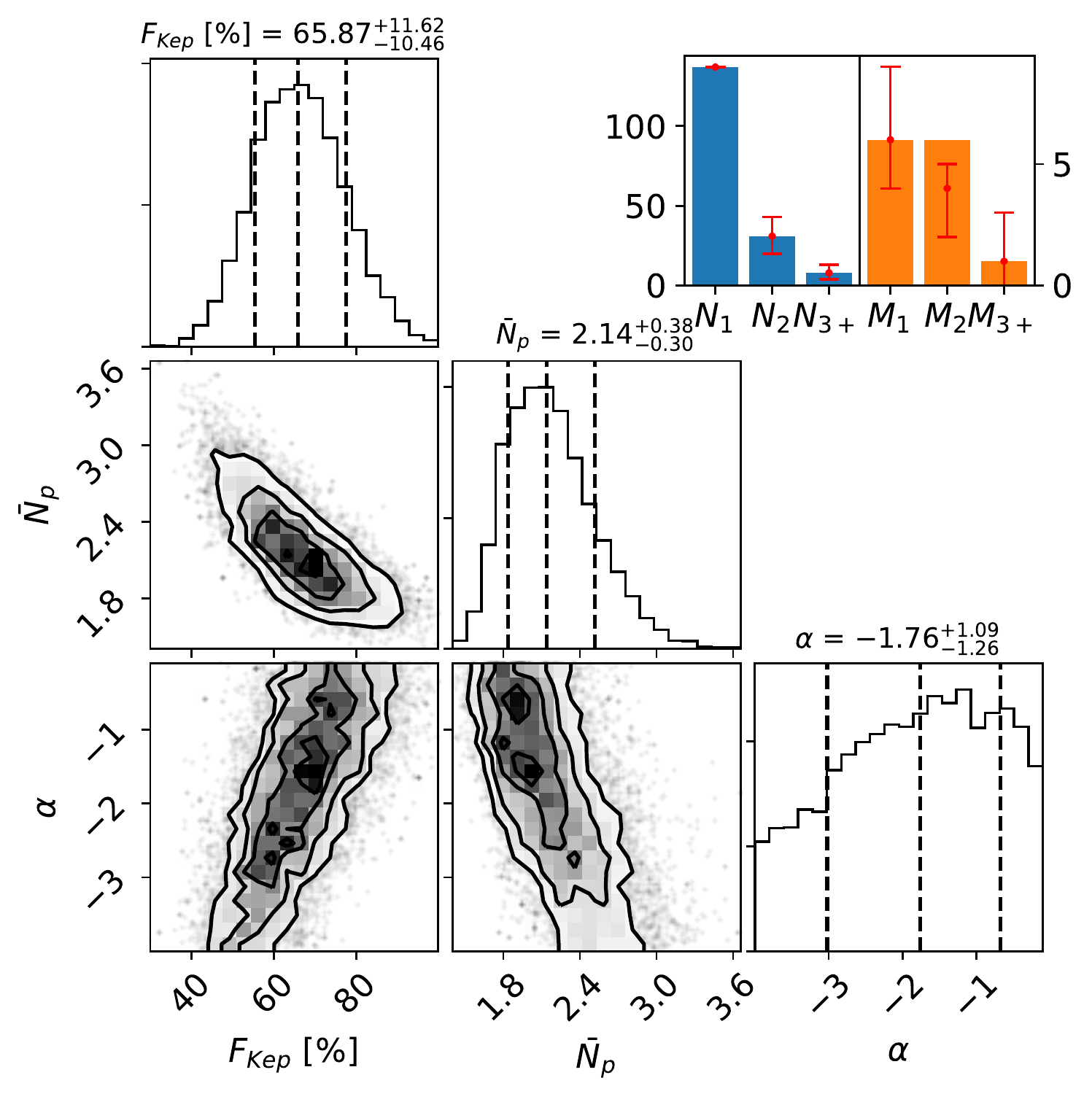}{0.5\textwidth}{Bin 6 (5876-6009 K)}
          }
\gridline{\fig{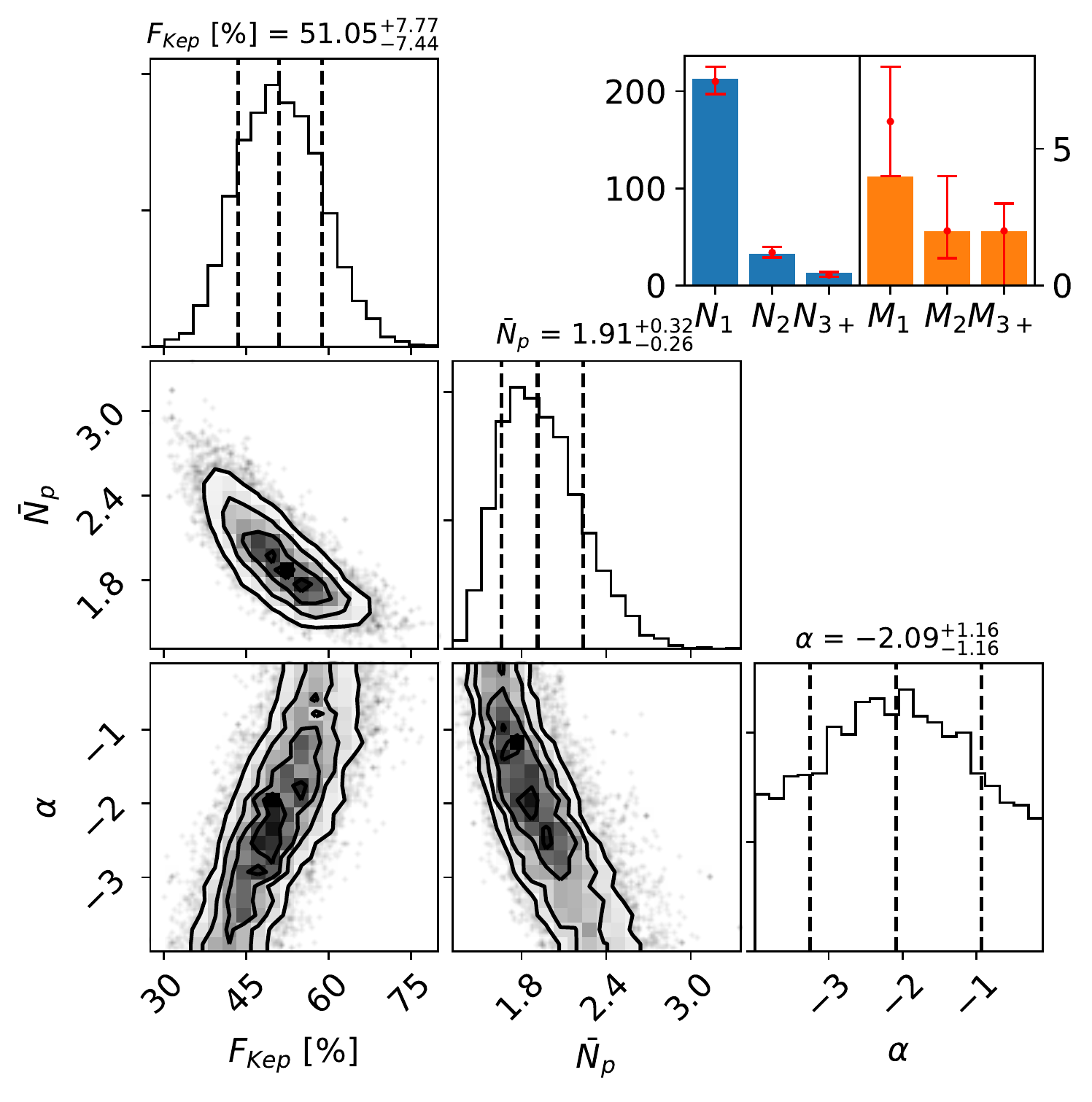}{0.5\textwidth}{Bin 7 (6009-6271 K)}
          \fig{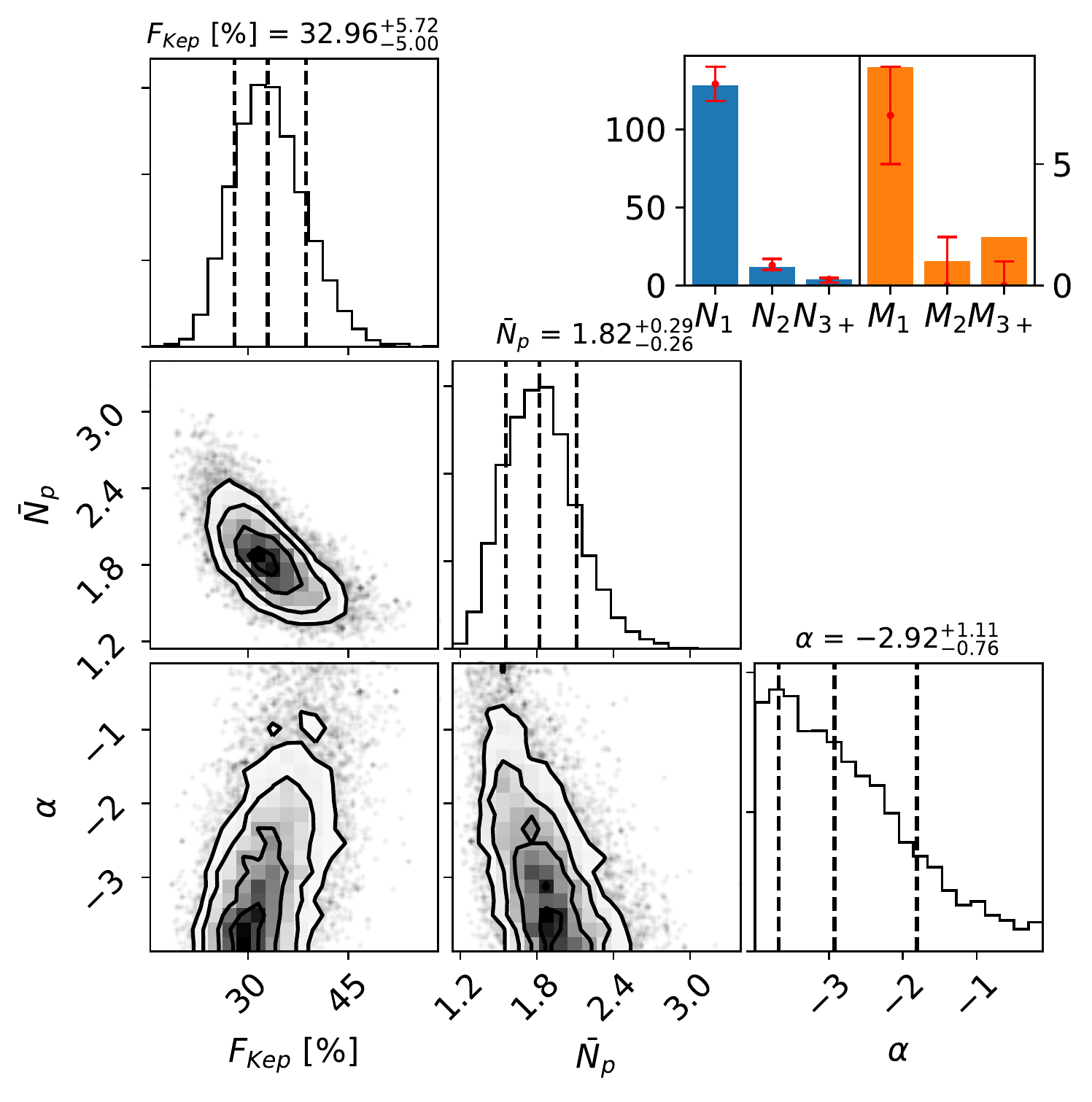}{0.5\textwidth}{Bin 8 (6271-7500 K)}
          }
\caption{Similar to Figure \ref{fig:fig_g1}, here are the MCMC results of the \teff\, bins 5-8. \label{fig:fig_os_m}}
\end{figure*}

In this section, we show more detailed results of our MCMC fitting results.
In Figure \ref{fig:fig_os_n} and Figure \ref{fig:fig_os_m}, we show the posterior distributions of \fkep, \npbar, and $\alpha$ (i.e., the MCMC corner plots) as well as the posteriors of multiplicity and TTV functions, and their comparisons to the observations for all the eight \teff\, bins.
The observed multiplicity functions and TTV functions fall right in the 1$\sigma$ range of the MCMC posteriors.
As we can see from the corner plots, both of \fkep\ and \npbar\ are constrained to Gaussian-like distributions, and they are anticorrelated.
Such an anticorrelation is not unexpected, because it generally reflects the fact that increase/decrease in \fkep \ can somewhat compensate for the decrease/increase in \npbar\ to yield a given number of tranets.
As for $\alpha$, which is greatly affected by TTV function, each bin gives slightly different results. 
For bins (e.g., bins 1, 3, and 8) which have relatively high $M_1$ and low $M_2$ and $M_{3+}$, $\alpha$ is constrained to $<-2$ with 1 $\sigma$ confidence. 
For bins with relatively lower $M_1$ and relatively higher $M_2$ or $M_{3+}$, $\alpha$ becomes larger, like bins 6 and 7.
Due to the small TTV sample size of in each bin, the constraint on $\alpha$ is generally loose with relatively large error bars. 
Nevertheless, for most bins, our results are largely consistent with that of \citet{Zhu_2018_ApJ_860_101Z}, namely orbital inclination is a steep falling function of planet multiplicity (Equation \ref{eq:inclination}), with the inclination index of $\alpha \sim < -2$.

\begin{figure*}[ht!]
\centering
\includegraphics[width=.4\textwidth]{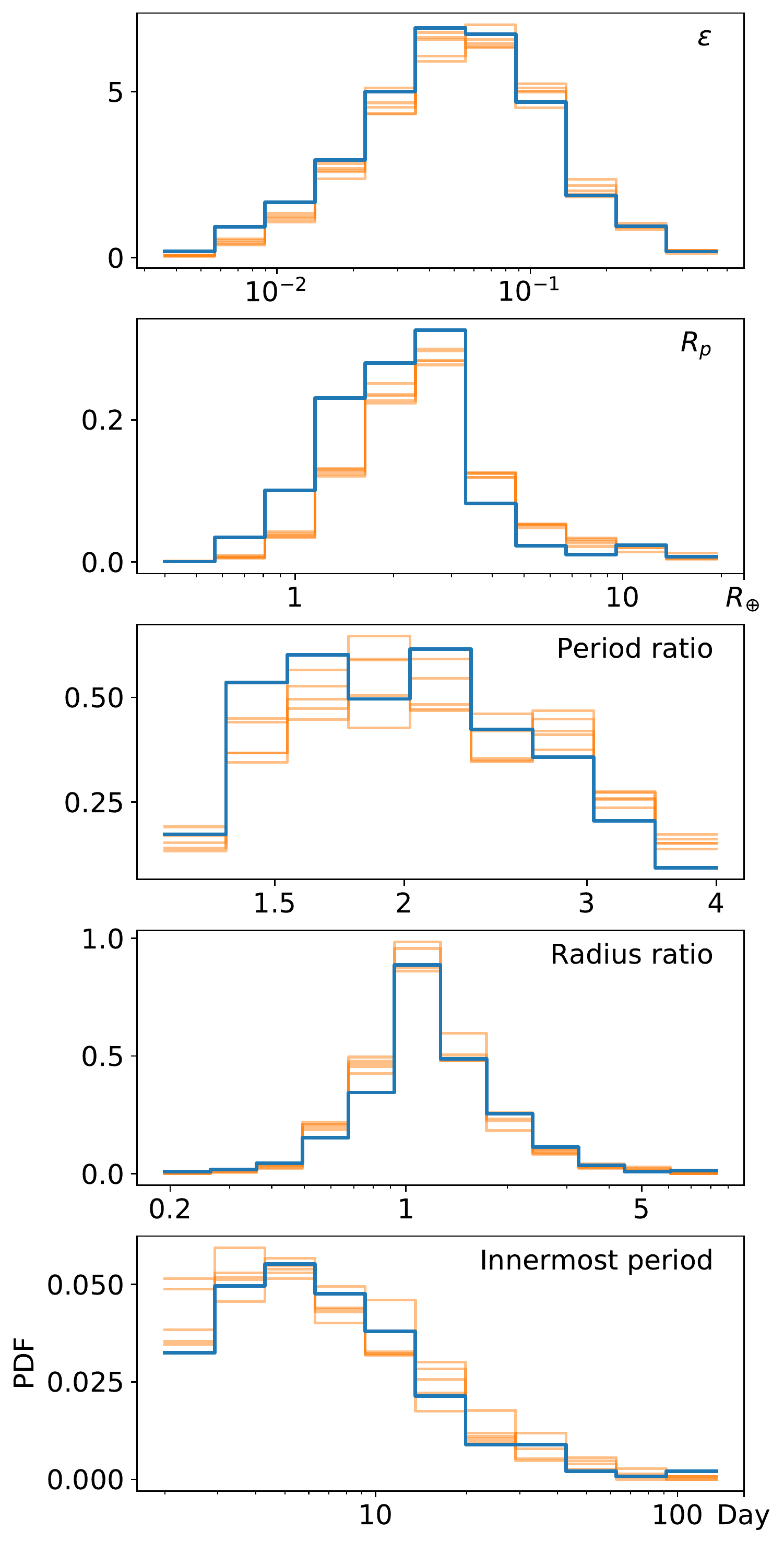}
\caption{Observed and simulated distributions of $\epsilon$, $R_p$, period ratio, radius ratio, and period of the innermost planet. The blue lines show distributions from observations, and the orange lines are results from five sets of simulations with \fkep, \npbar, and $\alpha$ taken from the MCMC posterior distributions.
\label{fig:fig_5_compare}}
\end{figure*}

We also checked whether our model could reproduce some general properties of the \kepler\ planet sample.
In Figure \ref{fig:fig_5_compare}, we compare the observed distributions of $\epsilon$, planet radii, prs, rrs, and the innermost orbital periods to those from simulations based on our MCMC posterior parameters. 
As can be seen, our model largely reproduces these distribution properties, though we note that the modeled distributions of radii and prs shift to larger values somewhat.
These subtle differences are not unexpected because our model is still not perfect and some substructures of planetary properties have not been considered.
For example, the period dependence of the radius gap \citep{Fulton.2017AJ....154..109F} is not taken into account in our model because we assign $\epsilon$ and radius separately, namely, they are treated as independent of each other. 
A sophisticated model that takes into account all the overall properties and various substructures of the observed sample is not trivial.
Although such a sophisticated model in principle can further improve the fit to observations (e.g., Figure \ref{fig:fig_5_compare}), it would not significantly change our main results (see also in Appendix \ref{sec:different_bins} and Figures \ref{fig:fig_fkep}-\ref{fig:fig_alpha}).

\section{Further Check of the \fkep--\teff, \npbar--\teff, and $\alpha$--\teff\ Trends}
\label{sec:different_bins}
\begin{figure*}[ht!]
\centering
\includegraphics[width=.6\textwidth]{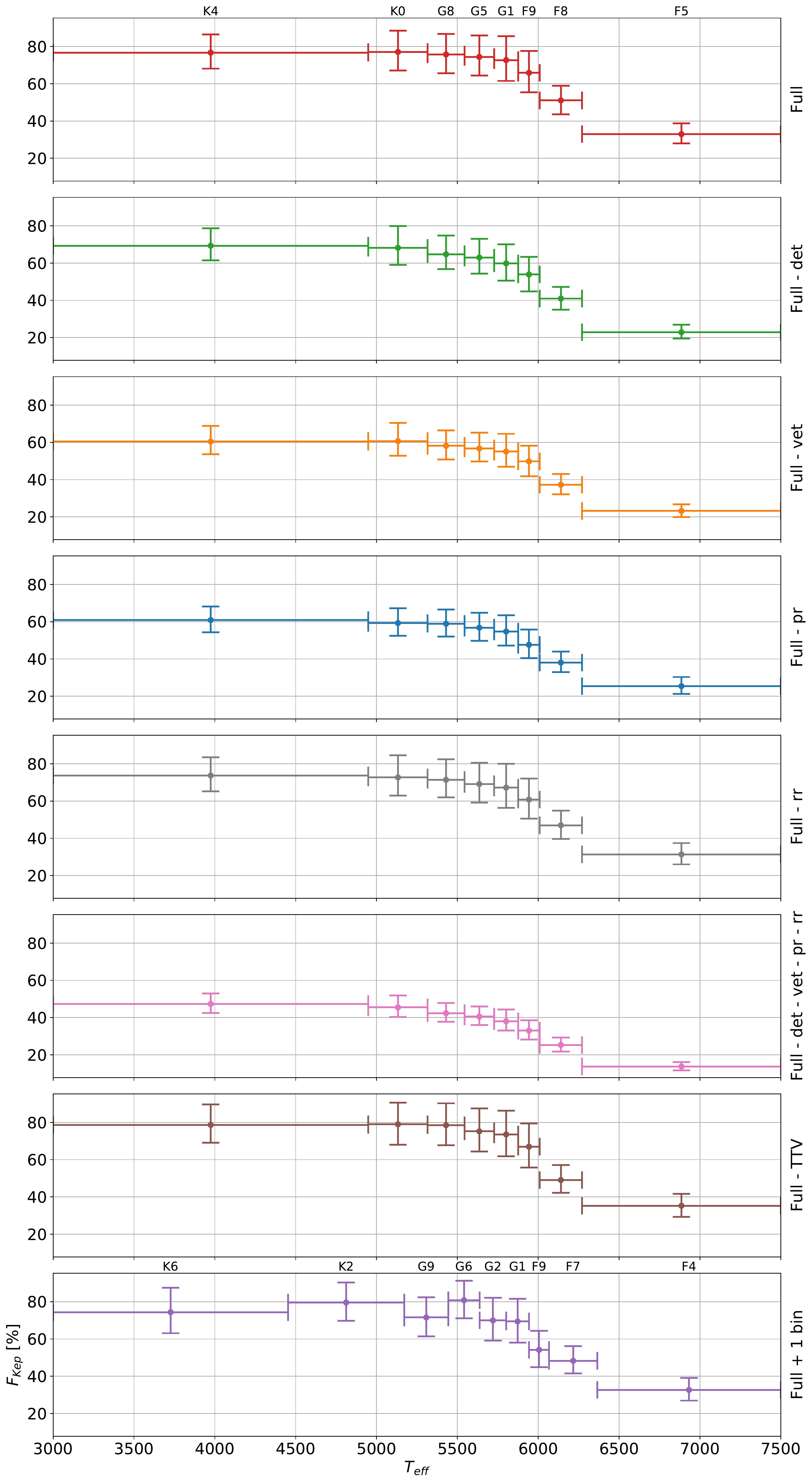}
\caption{Fraction of stars with \kepler\ planets (\fkep) as a function of stellar effective temperature (\teff) in different models. The top panel shows the result of our full model. Other panels show the results of different modified models with some ingredients being removed with respect to the full model, including a model without detection efficiency correction (Full-det), a model without vetting efficiency correction (Full-vet), a model without period ratio adjustment (Full-pr), a model without radius ratio adjustment (Full-rr), a model without detection and vetting efficiency correction and without period ratio and radius ratio adjustment (Full-pr-rr-det-vet), a model without TTV fitting (Full-TTV), and a model with one additional bin (Full+1 bin). See the text in Appendix \ref{sec:different_bins}.
\label{fig:fig_fkep}}
\end{figure*}

\begin{figure*}[ht!]
\centering
\includegraphics[width=.6\textwidth]{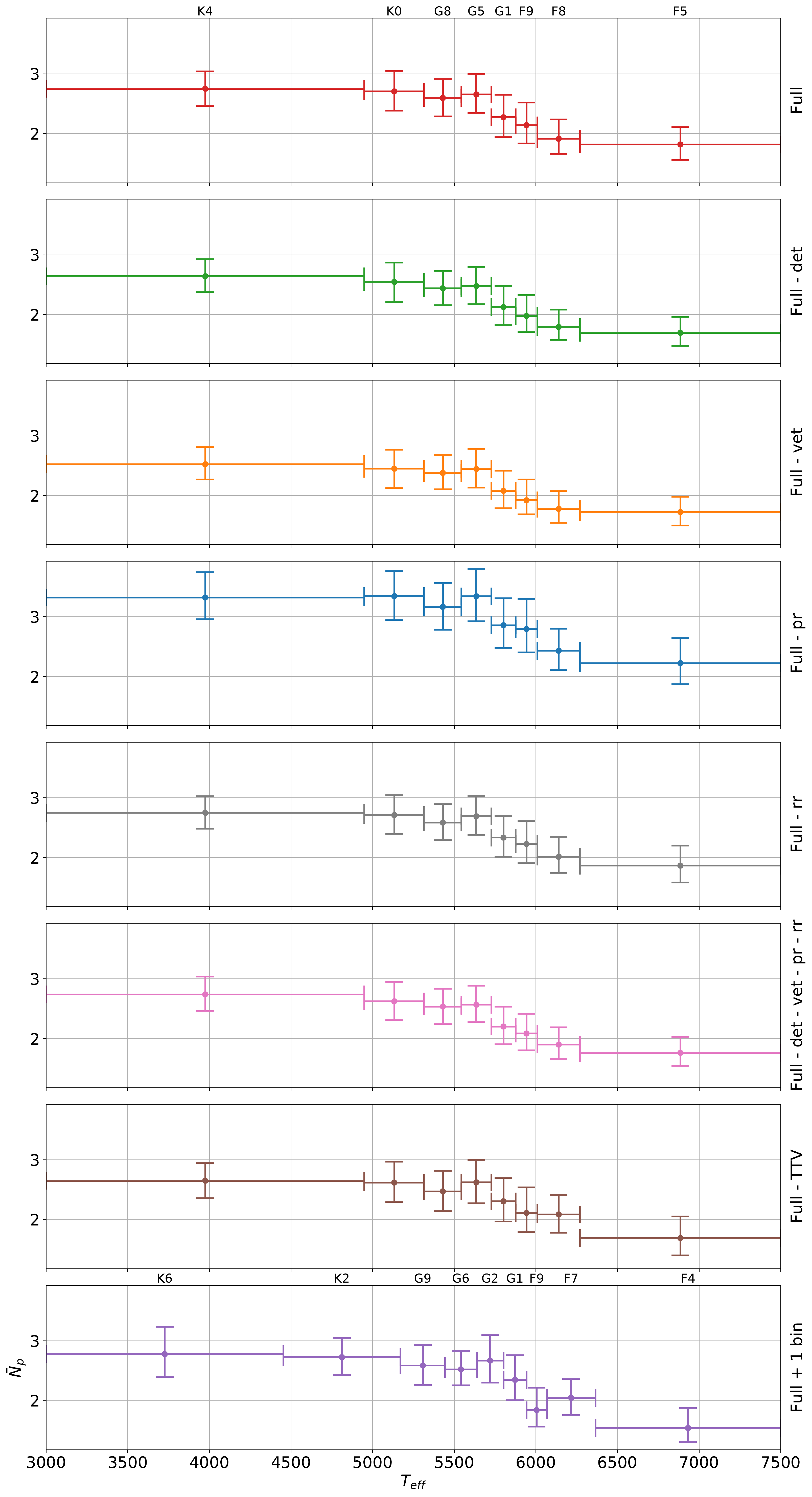}
\caption{Similar to Figure \ref{fig:fig_fkep}, but here are average planet multiplicity (\npbar) as a function of stellar effective temperature (\teff) in different models.
\label{fig:fig_npbar}}
\end{figure*}

\begin{figure*}[ht!]
\centering
\includegraphics[width=.6\textwidth]{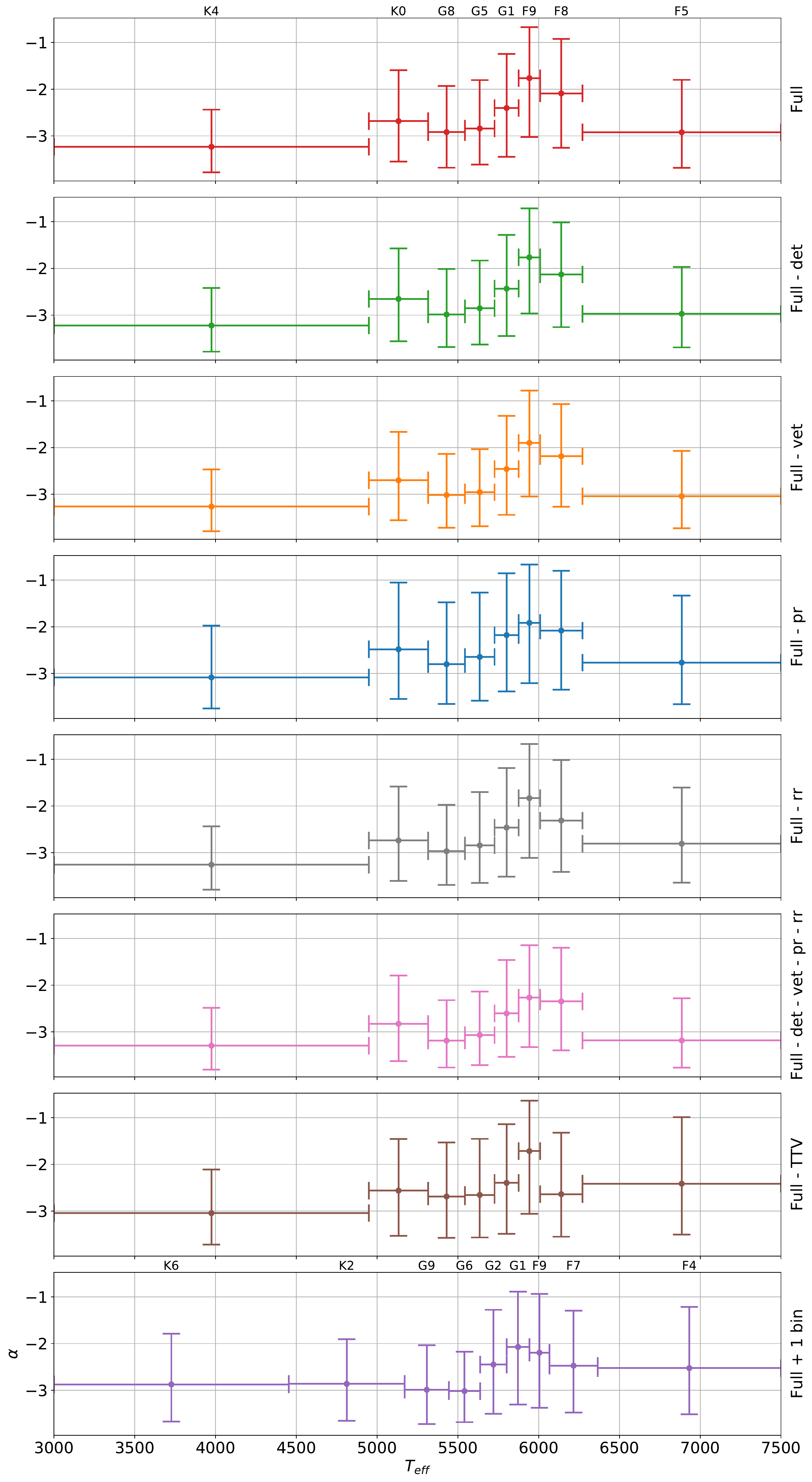}
\caption{Similar to Figure \ref{fig:fig_fkep}, but here are inclination slope index ($\alpha$) as a function of stellar effective temperature (\teff) in different models.
\label{fig:fig_alpha}}
\end{figure*}

As discussed in section \ref{sub:compare_to_zhu} and shown in Figure \ref{fig:fig_bin5}, removing some model ingredients can change the values of \fkep\ and \npbar. In this section, we further check how these model ingredients affect the \fkep-\teff\ (Figure \ref{fig:fig_fkep}), \npbar-\teff\ (Figure \ref{fig:fig_npbar}) and $\alpha$-\teff\ (Figure \ref{fig:fig_alpha}) trends.
We also consider a one additional bin method to test the bin effect. 
Specifically, we increase the number of total bins to 11 but reduce the bin sizes of the first and the last ones to 5000 and 6000 stars, respectively, and then merge the last four bins into two. 
The bin boundaries are (3000, 4453, 5171, 5444, 5638, 5802, 5943, 6067, 6365, and 7500 K).
Such a one additional method is essentially the same as the nominal one but with the bin boundaries shifted by about a half of bin width.
As can be seen, all results show similar trends, i.e., \fkep\ drops by about $\sim$40\%, and \npbar\ drops by $\sim$1 as \teff\ increases from 3000 to 7500 K with the transition occurring mainly between 5500 and 6000 K, and $\alpha$ generally varies between -3 and -2.
Different model ingredients do affect the results, but they only change the normalization factor not the relative degree of trends (e.g., Figure \ref{fig:fig_fkep}).

\section{Observed $\epsilon$, Radius, and Pr Distributions of Different \teff\, Bins}
\begin{figure*}[ht!]
\centering
\includegraphics[width=0.8\textwidth]{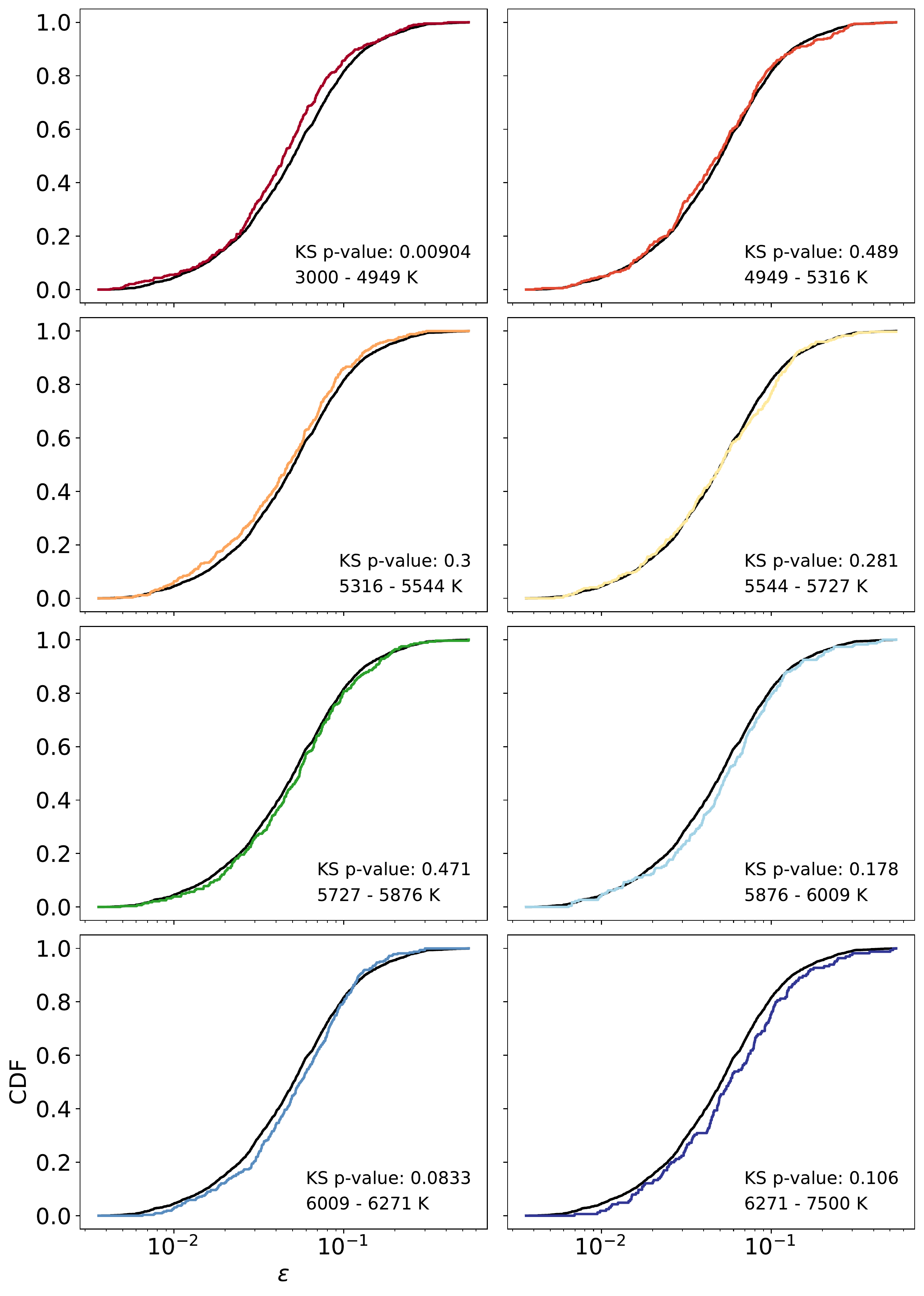}
\caption{Cumulative distribution functions (CDF) of the observed $\epsilon$ in the eight bins, and comparison to the whole sample (black line). In the lower right corner of each panel, we print the $p$-value of the KS test of the two distributions as well as the \teff\, range. As can be seen, $\epsilon$ distributions in most bins,  except for the first bin, are not significantly different from that of the whole sample.  
\label{fig:fig_cdf_eps}}
\end{figure*}

\begin{figure*}[ht!]
\centering
\includegraphics[width=0.8\textwidth]{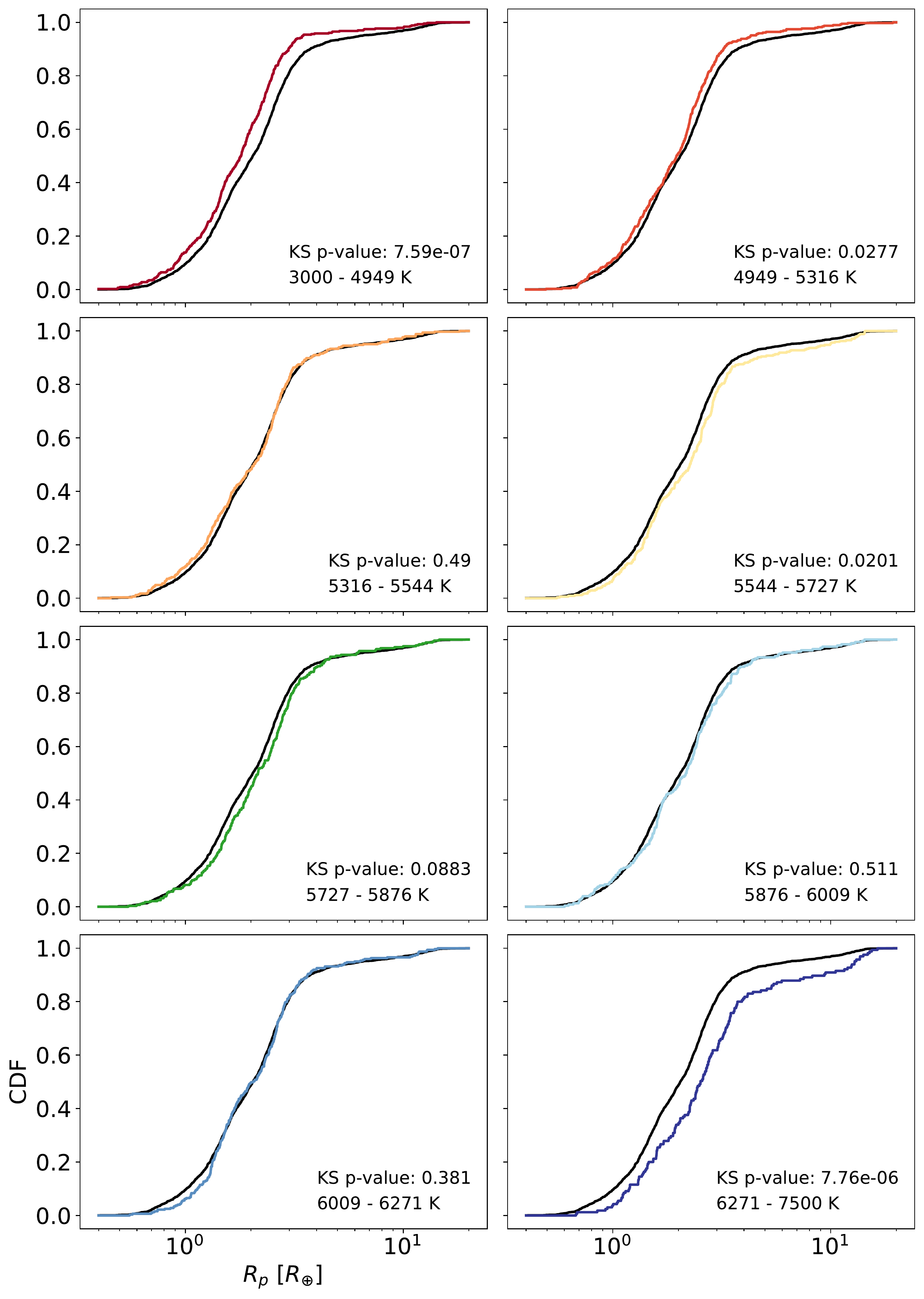}
\caption{Similar to Figure \ref{fig:fig_cdf_eps}, but with the CDF of observed radii in the eight bins shown. 
As can be seen, radius distributions in most bins,  except for the first and last bins, are not significantly different from that of the whole sample.  
\label{fig:fig_cdf_rp}}
\end{figure*}

\begin{figure*}[ht!]
\centering
\includegraphics[width=0.8\textwidth]{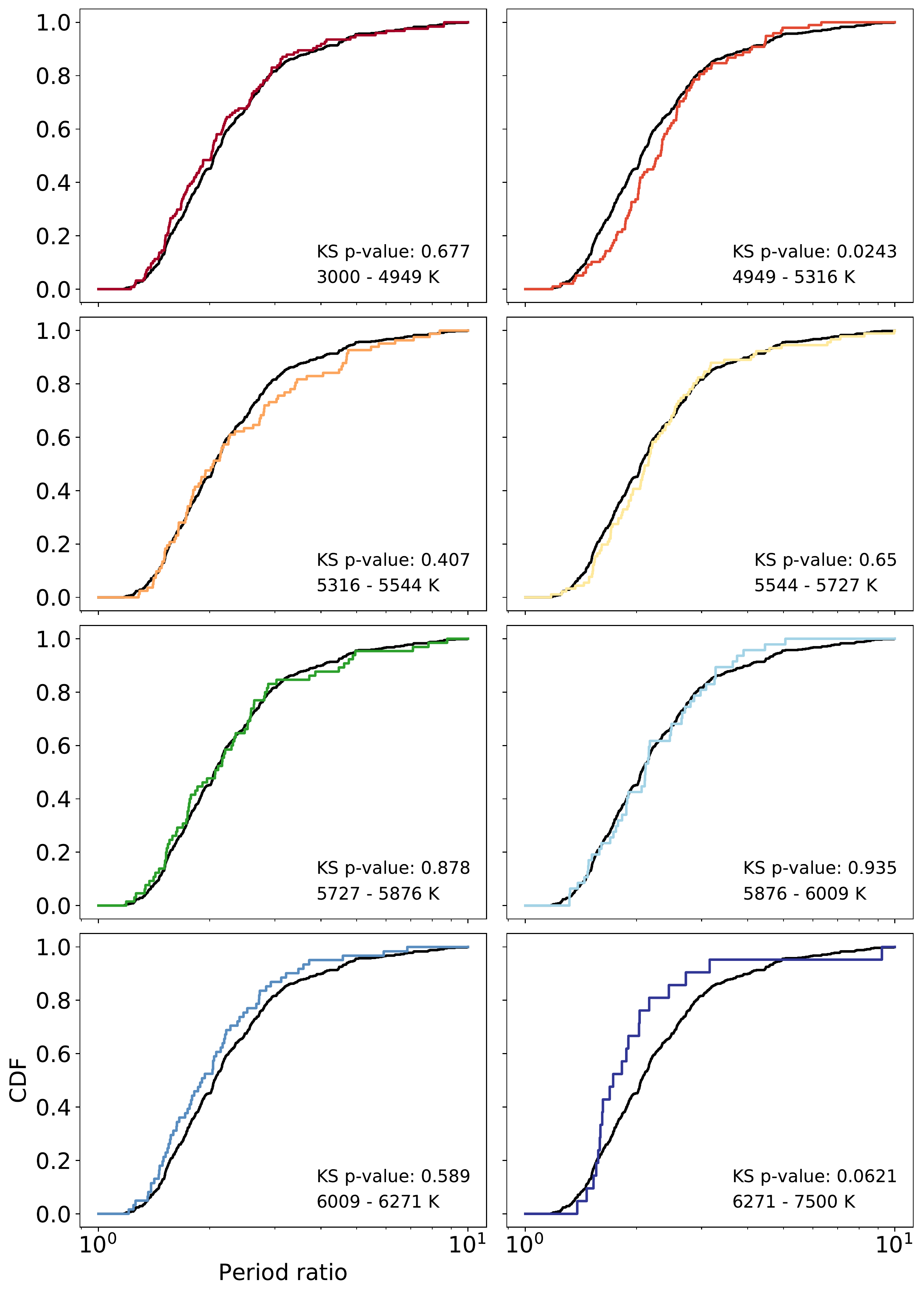}
\caption{Similar to Figure \ref{fig:fig_cdf_eps}, but with the CDF of observed period ratios in the eight bins shown. As can be seen, period ratio distributions in most bins are not significantly different from that of the whole sample.  
\label{fig:fig_cdf_pr}}
\end{figure*}

In our model, we adopt the same $\epsilon$, radius, and pr distributions from the whole sample to generate tranets for all the eight bins. 
Indeed, as we found in Figures \ref{fig:fig_cdf_eps}-\ref{fig:fig_cdf_pr}, these parameter distributions in most bins are not significantly different from those of the whole sample.
For those bins (a few) with significant different distributions in $\epsilon$, radius, and pr, we performed tests by adopting the $\epsilon$, radius, and pr distributions based on their own. 
The results only change slightly, and have little impact on the \fkep--\teff, \npbar--\teff, and $\alpha$--\teff\ trends.

\bibliographystyle{aasjournal}
\bibliography{yjylib}

%

\end{document}